\documentclass[preprint,12pt]{elsarticle}

\usepackage{amssymb}
\usepackage{amsthm,amsmath,amssymb}
\usepackage{mathrsfs}
\usepackage[section]{placeins}
\usepackage{subfigure}
\usepackage{booktabs}
\usepackage{multirow}
\usepackage{float}
\usepackage[dvipsnames]{xcolor}
\usepackage{algorithm}
\usepackage{algorithmic}

\DeclareMathOperator*{\argmin}{arg\,min}

\usepackage{lineno}

\journal{Neurocomputing}

\begin{document}

\begin{frontmatter}



\title{Swin Transformer for Fast MRI}


\author[label1,label2]{Jiahao Huang\corref{cor1}}
\ead{j.huang21@imperial.ac.uk}
\author[label1]{Yingying Fang}
\author[label1,label3]{Yinzhe Wu}
\author[label1,label3]{Huanjun Wu}
\author[label4]{Zhifan Gao}
\author[label5]{Yang Li}
\author[label6,label7]{Javier Del Ser}
\author[label8]{Jun Xia}
\author[label1,label2]{Guang Yang\corref{cor1}}
\cortext[cor1]{Corresponding authors.}
\ead{g.yang@imperial.ac.uk}

\address[label1]{National Heart and Lung Institute, Imperial College London, London, SW7 2AZ, United Kingdom}
\address[label2]{Cardiovascular Research Centre, Royal Brompton Hospital, London, SW3 6NP, United Kingdom}
\address[label3]{Department of Bioengineering, Imperial College London, London, SW7 2AZ, United Kingdom}
\address[label4]{School of Biomedical Engineering, Sun Yat-sen University, Guangzhou, 510275, China}
\address[label5]{School of Automation Sciences and Electrical Engineering, Beihang University, Beijing, 100190, China}
\address[label6]{Department of Communications Engineering, University of the Basque Country UPV/EHU, Bilbao, 48013, Spain}
\address[label7]{TECNALIA, Basque Research and Technology Alliance (BRTA), Derio, 48160, Spain}
\address[label8]{Department of Radiology, Shenzhen Second People’s Hospital, The First Affiliated Hospital of Shenzhen University Health Science Center, Shenzhen, 518037, China}

\begin{abstract}
Magnetic resonance imaging (MRI) is an important non-invasive clinical tool that can produce high-resolution and reproducible images.
However, a long scanning time is required for high-quality MR images, which leads to exhaustion and discomfort of patients, inducing more artefacts due to voluntary movements of the patients and involuntary physiological movements. 
To accelerate the scanning process, methods by \textit{k}-space undersampling and deep learning based reconstruction have been popularised. 
This work introduced SwinMR, a \textcolor{black}{novel} Swin transformer based method for fast MRI reconstruction. The whole network consisted of an input module (IM), a feature extraction module (FEM) and an output module (OM). The IM and OM were 2D convolutional layers and the FEM was composed of a cascaded of residual Swin transformer blocks (RSTBs) and 2D convolutional layers. The RSTB consisted of a series of Swin transformer layers (STLs). The shifted windows multi-head self-attention (W-MSA/SW-MSA) of STL was performed in shifted windows rather than the multi-head self-attention (MSA) of the original transformer in the whole image space.
A novel multi-channel loss was proposed by using the sensitivity maps, which was proved to reserve more textures and details.
We performed a series of comparative studies and ablation studies in the Calgary-Campinas public brain MR dataset and conducted a downstream segmentation experiment in the Multi-modal Brain Tumour Segmentation Challenge 2017 dataset. 
The results demonstrate our SwinMR achieved high-quality reconstruction compared with other benchmark methods, and it shows great robustness with different undersampling masks, under noise interruption and on different datasets. The code is publicly available at https://github.com/ayanglab/SwinMR.
\end{abstract}


% \begin{highlights}
% \end{highlights}

\begin{keyword}
MRI reconstruction \sep transformer \sep compressed sensing \sep parallel imaging


\end{keyword}

\end{frontmatter}

%\linenumbers

\section{Introduction}

% MRI
Magnetic resonance imaging (MRI) is an important non-invasive imaging technique, which enables excellent assessments of structural and functional conditions with no radiation in a reproducible manner. Basically, MRI is aimed to reconstruct the images from the observed signals whose degradation process can be formulated as follows: 

\begin{align}\label{formula:mri_forword}
y = \mathcal{F}x + n,
\end{align}

\noindent where $x,y \in \mathbb{C}^{N}$ are the vectors denoting the latent image to reconstruct in the image domain and the observed measurements in \textit{k}-space, $\mathcal{F} \in \mathbb{C}^{N \times N}$ is the \textcolor{black}{two-dimensional (2D)} discrete Fourier transform (DFT) and $n$ is the noise inevitably appearing in the signal acquisition process.

However, acquiring the full measurements of $y$ to construct a high-quality MR image $x$ is highly time-consuming. Moreover, the long scanning time will bring about the artefacts arising from the voluntary movements of the patients and involuntary physiological movements~\citep{Zbontar2018}. 
In order to mitigate the long acquisition time of MRI as well as alleviate the aliasing artefacts, a range of methods has been developed for accelerating MRI to obtain accurate reconstructions. 
Traditionally, \textcolor{black}{gradient refocusing~\citep{Stehling1991} and multiple-radio frequency mediated~\citep{Hennig1986} approaches were proposed.} Under constraints of the Nyquist-Shannon sampling theorem, they did reduce the scanning time although by only a limited factor. With the development of the parallel imaging (PI) and the compressed sensing (CS), the fast MRI based on these two theories attracted much research and advancements.

Parallel imaging was introduced to take advantage of spatial sensitivity distribution derived from an array of carefully distributed receiver surface coils, to reduce the measurement from each coil, alleviating the need of enhancing gradient performance and hence reducing the acquisition time~\citep{Blaimer2004}. 
The undersampled \textit{k}-space signal using \textcolor{black}{PI-MRI} can be represented by a general model as:

\begin{align}\label{formula:pi_mri_forword}
y^q = \mathcal{F}_{u}(\mathcal{S}^q \otimes x) + n^q, 
\quad q = 1,...S,
\end{align}

\noindent \textcolor{black}{where $\mathcal{S}^q$ and $n^q$ are the sensitivity map and inevitable noise of the $q^{\text{th}}$ coil ($S$ coils in total). $\otimes$ denotes the pixel-wise multiplication. $\mathcal{F}_{u} \in \mathbb{C}^{M \times N}$ is the undersampled 2D DFT matrix with $M\ll N$ to reduce the measurements of each $y^q$. With $S$ coils applied parallelly, one can obtain $y^{1}, ..., y^{S}$ simultaneously to reconstruct the latent image $x$.} To reconstruct these PI acquired images, great progress in developing PI reconstruction techniques has taken place, proposing popular methods such as the simultaneous acquisition of spatial harmonic (SMASH)~\citep{Sodickson1997}, sensitivity encoding (SENSE)~\citep{Pruessmann1999} and generalized auto-calibrating partially parallel acquisition (GRAPPA)~\citep{Griswold2002}. 

% CS
The invention of CS theory~\citep{Donoho2006} further advanced the sampling efficiency of MRI. 
The CS-MRI \textcolor{black}{utilises} the non-linear methodology and sparse transformation to reconstruct the latent image from only a small portion of \textit{k}-space measurement under a much smaller downsampling rate than the Nyquist rate. The general problem of MRI using the CS-MRI is to find the minimiser image to the following problem:

\begin{align}\label{formula:cs_mri_1}
\argmin_{x} \mid\mid\Phi x\mid\mid_{1}, 
\quad \text{s.t. } y=\mathcal{F}_{u} x,
\end{align}

\noindent where $\Phi$ is the sparsifying transformation, $\mathcal{F}_{u} \in \mathbb{C}^{M \times N}$ is undersampled 2D DFT with $M\ll N$, and $y \in \mathbb{C}^{M}$ is the observed undersampled measurements in \textit{k}-space. 
A range of non-linear reconstruction methods has demonstrated success in resolving this, including some fixed sparsifying methods such as \textcolor{black}{total variation}~\citep{Block2007}, curvelets~\citep{Beladgham2008} and double-density complex wavelet~\citep{Zhu2013}, and a few adaptive sparsifying models taking the advantage of dictionary learning~\citep{Ravishankar2011}.
While both CS-MRI and PI-MRI can significantly reduce the required number of measurements in \textit{k}-space, the iterative algorithms are required to derive the image however prolong the reconstruction time and hence cause concerns when \textcolor{black}{transferred} for actual clinical uses. 

% CNN
\textcolor{black}{As a modern popular method for general image analysis, deep learning has been very successful by exploiting the non-linear and complex natures of the network with supervised or unsupervised learning, and widely applied in medical image research ~\citep{Zeng2019,Zeng2021,Wu2022,Chen2022,Bakator2018}.}
Convolutional neural networks (CNNs) as a special type of deep learning networks enable enhanced latent feature extraction by their very deep hierarchical structure. CNN has demonstrated its superiority in multiple tasks, including detection~\citep{Girshick2014}, classification~\citep{Szegedy2015}, segmentation~\citep{Long2015} and super-resolution~\citep{Dong2016}. 
Wang et al.~\citep{Wang2016} became the pioneer to take advantage of CNNs by extracting latent correlations between undersampled and fully sampled \textit{k}-space data for MRI reconstruction. 
Yang et al.~\citep{Yang2016} further improved the network structures by re-applying the alternating direction method of multipliers (ADMM), which was originally used for \textcolor{black}{CS-MRI} reconstruction methods. 
A cascaded structure was developed by Schlemper et al.~\citep{Schlemper2018} for the more targeted reconstruction of dynamic sequences in cardiac MRI. 
To enable further latent mapping in the reconstruction model, Zhu et al.~\citep{Zhu2018} developed a novel framework to provide more dense mapping through domains via its proposed automated transform by manifold approximation.

\begin{figure}[H]
    \centering
    \includegraphics[width=5in]{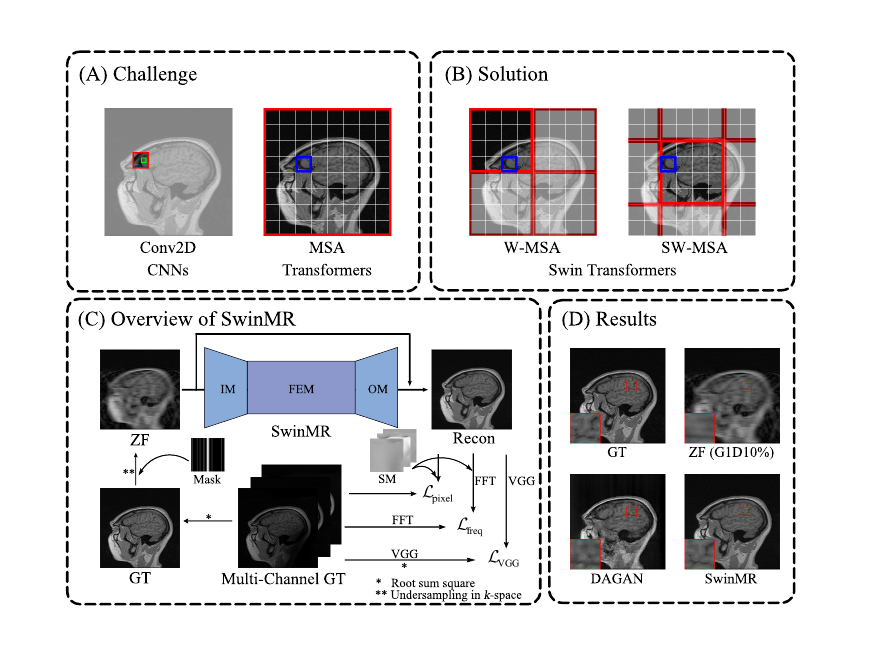}
    \caption{
    Overview of the proposed SwinMR. 
    (A) and (B) are the schematic diagrams of the receptive field for 2D convolution (Conv2D), multi-head self-attention (MSA) and shifted windows based multi-head self-attention (W-MSA/SW-MSA). Conv2D is locally sensitive and lacks long-range dependency. \textcolor{black}{Compared with Conv2D, MSA and (S)W-MSA have larger receptive fields. MSA is performed in the whole image space, while W-MSA and SW-MSA are alternatingly used in Swin transformer~\citep{Liu2021}, and performed in shifted windows.} 
    (Red box: the receptive field of the operation; green box: the pixel; blue box: the patch in self-attention.) 
    (C) is the overview of SwinMR. 
    (D) shows the results of the proposed SwinMR compared with GT, ZF and another method DAGAN~\citep{Yang2018}. 
    (IM: the input module; FEM: the feature extraction module; OM: the output module. ZF: undersampled zero-filled MR images; Recon: reconstructed MR images; Multi-Channel GT: multi-channel ground truth MR images; GT: single-channel ground truth MR images; MASK: the undersampling mask; SM: sensitivity maps.)
    }
    \label{fig:FIG_Overview}
\end{figure}

% Limitation of CNNs
For a long time, CNNs have had a dominant position in the field of computer vision (CV) since convolutions are effective feature extractors. Most deep learning-based MRI reconstruction methods are based on CNNs, including the GAN-based model. As Figure~\ref{fig:FIG_Overview}(A) shows, the feature extraction of CNNs is based on convolution, which is locally sensitive and lacks long-range dependency. The receptive field of CNNs is limited by the convolutional kernel and the network depth. \textcolor{black}{Oversized convolutional kernel brings huge computational cost, and overly-deep network depth can cause gradient vanishing.} 

% Transformer
A novel structure, transformer, taking advantage of even deeper mapping, sequence-to-sequence model design~\citep{Sutskever2014} and adaptive self-attention setting~\citep{Vaswani2017,Parikh2016,Cheng2016,Matsoukas2021} with expanding receptive fields (Figure~\ref{fig:FIG_Overview}(A))~\citep{Parmar2018,Salimans2017} has been proposed recently and been popularised in natural language processing (NLP) initially~\citep{Qiu2020}. 
Then it has been applied to object detection~\citep{Carion2020}, image recognition~\citep{Dosovitskiy2020} and extended to super-resolution~\citep{Parmar2018} for general image analysis.

% Transformer in MRI
With its superior ability in image reconstruction and synthesis as demonstrated in natural images, we could see transformers applied in MRI in many different ways. 
For synthesis, it has greatly enhanced cross-modality image synthesis (PET-to-MR by directional encoder~\citep{Shin2020}, T1-to-T2 by a pyramid structure~\citep{Zhang2021}, and MR-to-CT and T1/T2/PD by a novel aggregated residual transformer block~\citep{Dalmaz2021}). 
Variants of the transformer also enabled improved performance in reconstruction and super-resolution tasks. 
It was first applied on the reconstruction of brain MR imaging~\citep{Korkmaz2021_1}. 
Korkmaz et al.~\citep{Korkmaz2021_2} developed an unsupervised adversarial method to alleviate the scarce training sample populations. 
To further improve the quality of imaging, Feng et al.~\citep{Feng2021_1} enabled an end-to-end joint reconstruction and super-resolution. 
Feng et al.~\citep{Feng2021_2} further advanced the model for these dual tasks by incorporating the model with task-specific novel cross-attention modules.

% Swin Transform & SwinIR 
However, the shift from NLP tasks to CV tasks leads to challenges: 
(1) Difference in scale: visual elements (e.g., pixels) in CV tasks tend to vary substantially in scale unlike language elements (e.g., word tokens) in NLP tasks. 
(2) Higher resolution: the resolution of pixels in images (or frames) tend to be much higher than words in sentences.~\citep{Liu2021} 
Therefore, it is a trade-off for less computational complexity to limit the scale of self-attention in a local window, as Figure~\ref{fig:FIG_Overview}(A) and (B) shows. \textbf{S}hifted \textbf{win}dows (Swin) transformer~\citep{Liu2021} replaced the traditional multi-head self-attention (MSA) by the shifted windows based multi-head self-attention (W-MSA/SW-MSA). \textcolor{black}{W-MSA and SW-MSA were alternatingly used in consecutive transformer layers, since if all attention operations are conducted in fixed windows, the cross-window relationship may be ignored.}
Based on the Swin transformer module, Liang et al.~\citep{Liang2021} proposed SwinIR for image restoration tasks.

% THIS WORK
In this work, we introduced the SwinMR, a novel parallel imaging coupled Swin transformer based model for fast \textcolor{black}{CS-MRI} reconstruction, as Figure~\ref{fig:FIG_Overview}(C) shows. The main contributions can be summarised as follows:

\begin{itemize}
\item[$\bullet$] 
A novel parallel imaging coupled Swin transformer-based model for fast MRI reconstruction was proposed, as Figure~\ref{fig:FIG_Overview}(C) shows.
\item[$\bullet$]
A novel multi-channel loss was proposed by using the sensitivity maps, which was proved to \textcolor{black}{preserve} more textures and details in the reconstruction results.
\item[$\bullet$] 
A series of ablation studies and comparison experiments were conducted. Experimental studies using different undersampling trajectories with various noises were performed to validate the robustness of our proposed SwinMR.
\item[$\bullet$] 
A downstream task experiment using a segmentation network was conducted. A pre-trained segmentation network was applied to test the segmentation score for reconstructed images. 
\end{itemize}

\section{Method}

\subsection{Classic Model-Based \textcolor{black}{CS-MRI} Reconstruction}

To recover better spatial information with less artefacts from the undersampled \textit{k}-space data, traditional CS-MRI methods usually consider solving the following optimisation problem:

\begin{align}\label{formula:cs_mri_2}
\min _{x} \frac{1}{2} \mid\mid \mathcal{F}_{u} x-y \mid\mid_{2}^{2}+ \lambda R(\Phi x),
\end{align}

\noindent where $\Phi$ is the sparsifying transform, e.g., discrete wavelet transform~\citep{qu2012undersampled}, gradient operator~\citep{Block2007, wu2017solving} and dictionary-based transform~\citep{ravishankar2010mr}. $R(\cdot)$ is the regularisation function imposed on the sparsity, e.g, $l_1$-norm and $l_0$-norm, and $\lambda$ is the weight parameter to balance the two terms. 
The solution of the above problem can be derived by the non-linear optimisation solvers such as gradient-based algorithms~\citep{lustig2007sparse} and variable splitting methods~\citep{yang2010fast, wang2014compressed}. Depending on the manually designed regularisation, some models may suffer from a long reconstruction time for better reconstruction quality. Additionally, the manually selected sparsifying transform $\Phi$ could also introduce undesirable artefacts, e.g., \textcolor{black}{total variation based} regularisation which is well-known for removing the noise and preserving the sharp edges can introduce staircase artefacts~\citep{Beladgham2008} and the tight wavelet frame transform increases the reconstruction efficiency but may lead to the blocky artefacts~\citep{cai2020data}.

\subsection{CNN-based Fast MRI Reconstruction}

To relieve the artefacts brought by the hand-crafting regularisation and the long reconstruction time of classic models, the \textcolor{black}{deep CNNs} which are well-known as the powerful features extractors, \textcolor{black}{were} firstly applied in the CS-MRI in~\citep{Wang2016}. In this work, a deep \textcolor{black}{CNN} was applied to learn the mapping from down-sampled reconstruction images to fully sampled reconstruction images directly.
Following that, several networks have been proposed to further improve the reconstruction quality.

Some works attempted to bridge the classic models with deep CNNs by mimicking the iterative algorithm in their network architectures. 
Deep ADMM Net~\citep{Yang2016} was firstly trained by unfolding the optimisation algorithm ADMM to derive the solution to the general model Equation (\ref{formula:cs_mri_2}) by network blocks.
% \begin{align}\label{formula:admm_mri}
% \min _{x} \frac{1}{2} \mid\mid \mathcal{F}_{u} x-y \mid\mid_{2}^{2}+ 
% \sum_{l=1}^{L} \lambda_{l} g\left(z_{l}\right),
% \text { s.t. } z_{l}=\Phi_{l} x, \quad \forall l \in\{1,2, \cdots, L\}.
% \end{align}
In~\citep{Schlemper2018}, the reconstruction of the deep CNN from lower-quality \textcolor{black}{images} was adopted as the prior information to approximate in a classic CS-model as follows:

\begin{align}\label{formula:cnn_mri}
\min_{x} \frac{1}{2} 
\mid\mid y - \mathcal{F}_u x\mid\mid^2_2
+ \lambda \mid\mid x - f_{\mathrm{CNN}}(x_u\mid\theta) \mid\mid^2_2,
\end{align}

\noindent where the solution of the above function was further adopted into the network architecture iteratively to improve the reconstruction result of $f_{\mathrm{CNN}}$ which takes the zero-filled reconstruction $x_u$ as the input. 

% GAN
On top of the CNNs, conditional generative adversarial networks (cGANs) exploited the advantages of deep learning further and proved to enhance the quality of the MR image reconstruction to a large extent~\citep{Yang2021, Lv2021_1}.
Such a competitive network introduced a two-player generator-discriminator training mechanism to competitively improve reconstruction performance by alternatingly optimising $\theta_{G}$ and $\theta_{D}$ of the generator $G$ and the discriminator $D$, in a general form as:

\begin{align}\label{formula:gan1}
\min_{\theta_{G}} \max_{\theta_{D}}
{\mathbb{E}}_{x \sim p_{\text{gt}}} 
\left[\log D_{\theta_{D}}(x)\right] +
{\mathbb{E}}_{{x_u \sim p_{\text{u}}}}\left[\log 
\left(1-D_{\theta_{D}}\left(G_{\theta_{G}}(x_u)\right)\right)\right],
\end{align}

\noindent where $G_{\theta_{G}}$ and $D_{\theta_{D}}$ denote the generator and the discriminator with parameters $\theta_{G}$ and $\theta_{D}$, respectively. $x$ and $x_u$ denote the ground truth MR images and undersampled zero-filled MR images with aliasing artefacts. After the training, the generator can yield the corresponding reconstruction from $x_u$ to reconstructed images $G_{\theta_{G}}(x_u)$.

Variants of generators and discriminators have been developed to cope with multiple flaws in the original architecture of GAN -- for improved generator~\citep{Shaul2020}, improved discriminator~\citep{Huang2021}, loss functions~\citep{Quan2018}, regularisation~\citep{Ma2021}, training stability by Wasserstein GAN~\citep{Arjovsky2017,Guo2020} and attention mechanism~\citep{Jiang2021}.
DAGAN~\citep{Yang2018}, by substituting the residual networks with \textcolor{black}{a modified} U-Net~\citep{Ronneberger2015}, combined the advantage of U-Net in latent information extraction with competitive training and pre-trained VGG based transfer learning.
Furthermore, PIDDGAN~\citep{Huang2021} considered edge information into their model and further enhance the edge information in the reconstruction, which are clinically important when interpreting MR images.
The utilisation of transfer learning improved the generalisability of a network trained with a small dataset~\citep{Lv2021_3}.

\textcolor{black}{CNN-based} MR reconstruction methods showed their superiority both on reconstruction quality and efficiency compared to classical MR reconstruction methods. 
However, the performance of those CNN-based methods was limited by the local sensitivity of the convolutional operation. Motivated by this limitation, we proposed a Swin transformer based MR reconstruction method SwinMR.

\subsection{SwinMR: Swin Transformer for MRI Reconstruction}

\subsubsection{Overall Architecture}

\begin{figure}[H]
    \centering
    \includegraphics[width=5in]{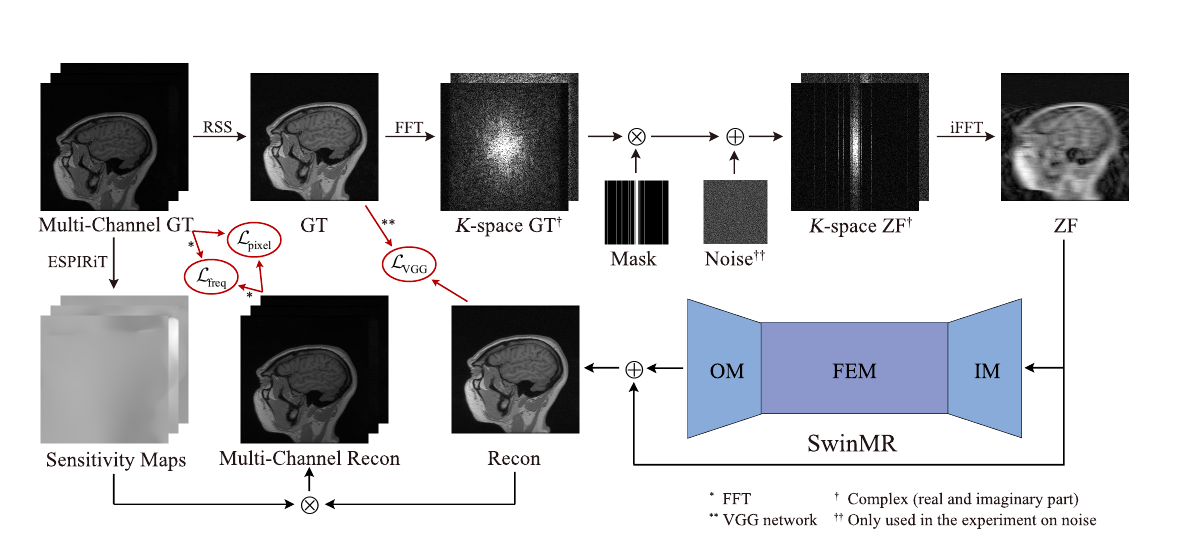}
    \caption{The dataflow of proposed SwinMR. 
    Root sum square (RSS) is applied to combine the multi-channel ground truth MR images (Muti-Channel GT) to single-channel ground truth MR images (GT). 
    Undersampling and noise interruption are performed in \textit{k}-space using fast Fourier transform (FFT) and inverse fast Fourier transform (iFFT) to convert the GT to undersampled zero-filled MR images (ZF) as the input of our proposed SwinMR. 
    Multi-channel reconstructed MR images (Muti-Channel Recon) are calculated by the pixel-wise multiplication of single-channel reconstructed MR images (Recon), which are the output of the proposed SwinMR, and sensitivity maps, which are estimated by ESPIRiT from the Multi-Channel GT.
    }
    \label{fig:FIG_DataFlow}
\end{figure}

The overall architecture is shown in Figure~\ref{fig:FIG_Overview}(C) and the data flow of SwinMR is shown in Figure~\ref{fig:FIG_DataFlow}. 
Root sum square (RSS) is applied to combine the multi-channel ground truth MR images $x^q$ to single-channel ground truth MR images $x$ ($q$ denotes the $q^{\text{th}}$ coil). Sensitivity maps $\mathcal{S}^q$ are estimated by ESPIRiT~\citep{Uecker2014} from multi-channel ground truth MR images $x^q$.
Undersampling and noise interruption are performed in \textit{k}-space using fast Fourier transform (FFT) and inverse fast Fourier transform (iFFT) (Gaussian noise is added in the noise experiments), which converts single-channel ground truth MR images $x$ to undersampled zero-filled MR images $x_u$.

The proposed SwinMR model can produce reconstructed MR images $\hat x_u$ from undersampled zero-filled MR images $x_u$, where the residual connection is applied to accelerate the convergence and stable the training processing. 
It can be expressed by

\begin{align}\label{formula:5}
\hat x_u = \text{SwinMR}(x_u \mid \theta) + x_u,
\end{align}

\noindent where the SwinMR network is parameterised by $\theta$.

\begin{figure}[H]
    \centering
    \includegraphics[width=5in]{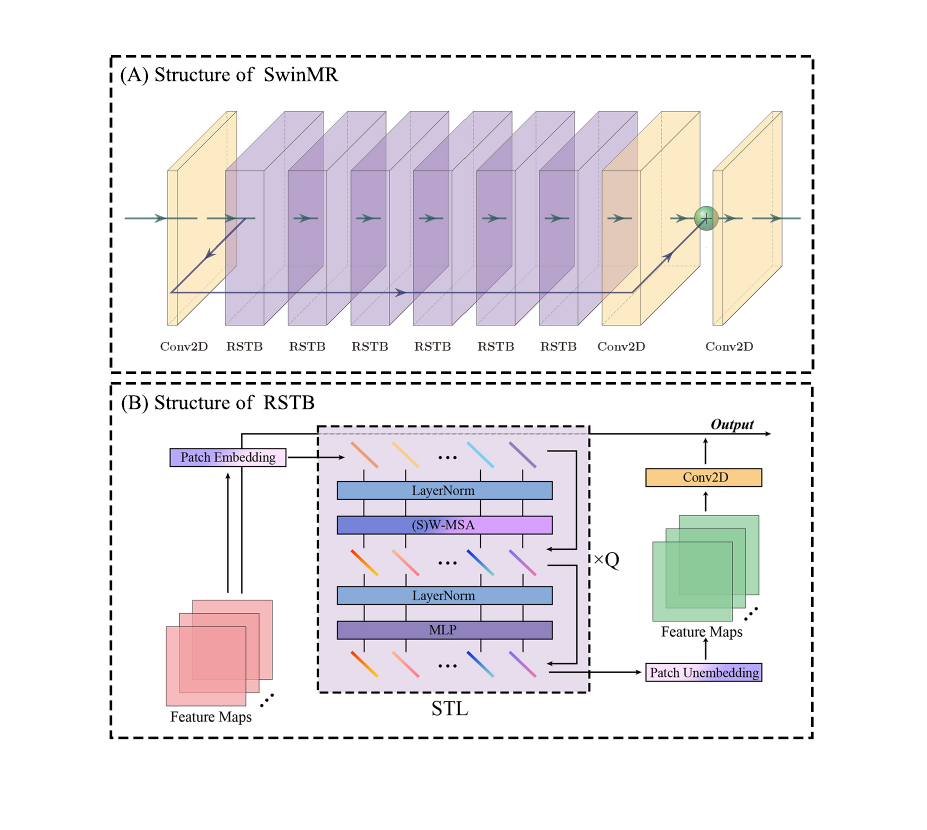}
    \caption{The structure of proposed SwinMR. 
    (A) shows the overall structure of SwinMR. In SwinMR architecture, two Conv2Ds are placed at the beginning and the ending. A cascade of RSTBs and a Conv2D with a residual connection are placed between the two Conv2Ds.
    (B) shows the structure of RSTB. The RSTB consists of a patch embedding operator, \textcolor{black}{$Q$} cascaded STLs, a patch unembedding operator, a Conv2D, and a residual connection between the input and output of RSTB. An STL consists of an LN, an (S)W-MSA, an LN and an MLP, with two residual connections. 
    (RSTB: the residual Swin transformer block; STL: the Swin transformer layer; Conv2D: the 2D convolutional layer; LN: the layer normalisation layer; \textcolor{black}{MLP: the multi-layer perceptron; (S)W-MSA: the (shifted) windows multi-head self-attention. W-MSA and SW-MSA are altinatively used in consecutive STLs.})
    }
    \label{fig:FIG_Structure}
\end{figure}

Figure~\ref{fig:FIG_Structure}(A) shows the structure of SwinMR, which is composed of an input module (IM), a feature extraction module (FEM) and an output module (OM). The IM and OM are at the beginning and the end of the whole structure, and the FEM is placed between the IM and OM with a residual connection. 
The structure can be expressed by

\begin{align}\label{formula:6}
& F_{\text{IM}} = \text{H}_{\text{IM}}(x_u),\\
& F_{\text{FEM}} = \text{H}_{\text{FEM}}(F_{\text{IM}}),\\
& F_{\text{OM}} = \text{H}_{\text{OM}}(F_{\text{FEM}}+F_{\text{IM}}),
\end{align}

\noindent where the $\text{H}_{\text{IM}}(\cdot)$, $\text{H}_{\text{FEM}}(\cdot)$ and $\text{H}_{\text{OM}}(\cdot)$ denote the IM, FEM and OM respectively. $F_{\text{IM}}$, $F_{\text{FEM}}$ and $F_{\text{OM}}$ denote the output of the IM, FEM and OM respectively.

\subsubsection{Input Module and Output Module}

The IM is used for early visual processing and mapping from the input image space to higher dimensional feature space for the following FEM. 
The IM applies a 2D convolutional layer (Conv2D) mapping $x_u \in \mathbb{R}^{H \times W \times 1}$ to $F_{\text{IM}} \in \mathbb{R}^{H \times W \times C}$.
In contrast, the OM is used to map the higher dimensional feature space to the output image space by a Conv2D mapping $F_{\text{FEM}} \in \mathbb{R}^{H \times W \times C}$ to $F_{\text{OM}} \in \mathbb{R}^{H \times W \times 1}$.

In the training stage, the input image is randomly cropped to a fixed size $H \times W$ ($H=W$). In the inference stage, $H$, $W$ denote the height and weight of the input image. Here we define $H$ (or $W$) as the patch number and $C$ as the channel number for the self-attention processing. 

\subsubsection{Feature Extraction Module}

The FEM is composed of a cascade of residual Swin transformer blocks (RSTBs) and a Conv2D at the end. It can be expressed as 

\begin{align}\label{formula:7}
& F_0 = F_{\text{IM}}, \\
& F_i = {\text{H}}_{\text{RSTB}_i} (F_{i-1}), \quad i = 1,2,...,P, \\
& F_{\text{FEM}} = {\text{H}}_{\text{CONV}}(F_P), 
\end{align}

\noindent where $F_{\text{IM}}$ and $F_{\text{FEM}}$ are the input and output of the FEM. ${\text{H}}_{\text{RSTB}_i}(\cdot)$ denotes the $i^{\text{th}}$ RSTB ($P$ RSTBs in total) in the FEM. ${\text{H}}_{\text{CONV}}(\cdot)$ denotes the Conv2D after a series of RSTBs.

Figure~\ref{fig:FIG_Structure}(B) shows the structure of the RSTB. An RSTB consists of $Q$ Swin transformer layers (STLs) and a Conv2D, and a residual connection is linked between the input and output of the RSTB. It can be expressed as 

\begin{align}\label{formula:8}
& F_{i,0} = {\text{H}}_{\text{Emb}_{i}}(F_{i-1}), \\
& F_{i,j} = {\text{H}}_{\text{STL}_{i,j}} (F_{i,j-1}), \quad j = 1,2,...,Q, \\
& F_{i} = {\text{H}}_{\text{CONV}_{i}}({\text{H}}_{\text{Unemb}_{i}}(F_{i,Q}) + F_{i-1}),
\end{align}

\noindent where ${\text{H}}_{\text{Emb}_{i}}(\cdot)$ is the patch embedding from $F_{i-1} \in \mathbb{R}^{H \times W \times C}$ to $F_{i,0} \in \mathbb{R}^{HW \times C}$, and ${\text{H}}_{\text{Unemb}_{i}}(\cdot)$ is the patch unembedding from $F_{i,Q} \in \mathbb{R}^{HW \times C}$ to $\mathbb{R}^{H \times W \times C}$.

${\text{H}}_{\text{STL}_{i,j}}(\cdot)$ and ${\text{H}}_{\text{CONV}_{i}}(\cdot)$ denote the $j^{\text{th}}$ STL and the Conv2D in the $i^{\text{th}}$ RSTB, respectively. 

\subsubsection{Swin Transformer Layer}

The whole process of the STL can be expressed as 

\begin{align}\label{formula:9}
&X^{\prime}={\text{H}}_{\text{(S)W-MSA}}({\text{H}}_{\text{LN}}(X))+X,\\
&X^{\prime\prime}={\text{H}}_{\text{MLP}}({\text{H}}_{\text{LN}}(X^{\prime}))+X^{\prime},
\end{align}

\noindent where $X$ and $X^{\prime\prime}$ are the input and output of the STL. ${\text{H}}_{\text{MLP}}(\cdot)$ and ${\text{H}}_{\text{LN}}(\cdot)$ denote the multilayer perceptron and the layer normalisation layer. \textcolor{black}{Windows multi-head self-attention (W-MSA) and shifted windows multi-head self-attention (SW-MSA) ${\text{H}}_{\text{(S)W-MSA}}(\cdot)$ are alternatingly applied in consecutive STLs.}

Spatial constraints are added in the Swin transformer layer compared to the original transformers. Figure~\ref{fig:FIG_Structure}(B) shows the W-MSA and the SW-MSA compared with the original MSA. Original MSA performs self-attention in the whole image space. Although the information of the entire picture is involved in each attention calculation, it aggravates computational costs and redundant connections. The computational complexity for the original MSA is as follows:

\begin{align}\label{formula:10}
\Omega({\text{H}}_{\text{MSA}})=4HWC^{2}+2(HW)^{2}C.
\end{align}

In Swin transformer layers, a $\mathbb{R}^{H \times W \times C}$ feature map are divided into $\frac{HW}{M^2}$ non-overlapped windows with the size of $M^2 \times C$. (S)W-MSA is calculated in each window, instead of the whole image space. The computational complexity for (S)W-MSA is as follows:

\begin{align}\label{formula:11}
\Omega({\text{H}}_{\text{(S)W-MSA}})=4HWC^{2}+2M^{2}HWC,
\end{align}

\noindent which is significantly reduced compared to the original MSA. However, if the separation of windows is fixed between \textcolor{black}{STLs}, the network will lose the link between different windows. Normal windows and shifted windows are alternatingly utilised in \textcolor{black}{consecutive STLs} to enable information communication from different windows.

(S)W-MSA for each non-overlap window $X$ can be expressed by

\begin{align}\label{formula:12}
Q=X P_{Q}, \quad K=X P_{K}, \quad V=X P_{V},
\end{align}

\noindent where the $P_{Q}$, $P_{K}$, $P_{V}$ are shared projection matrices over all the windows. The query $Q$, key $K$, value $V$ and learnable relative position encoding $B$ ($\mathbb{R}^{M^2 \times d}$) are used in the calculation of the self-attention mechanism in a local window, which can be expressed by

\begin{align}\label{formula:13}
\operatorname{Attention}(Q, K, V)=\operatorname{SoftMax}\left(Q K^{T} / \sqrt{d}+B\right) V.
\end{align}

Such self-attention mechanism calculations are performed for $h$ times and concatenated for (S)W-MSA. \textcolor{black}{The pseudo-code of STL and (S)W-MSA are shown in Algorithm~\ref{alg:stl} and Algorithm~\ref{alg:swmsa}.}

\begin{algorithm}[H]
	\renewcommand{\algorithmicrequire}{\textbf{Input:}}
	\renewcommand{\algorithmicensure}{\textbf{Output:}}
	\caption{Swin transformer layer (STL).} 
	\label{alg:stl} 
	\setlength{\tabcolsep}{4mm}{
    \begin{algorithmic}
        \REQUIRE $X, j, W_s, N_h$
        \STATE \textcolor{OliveGreen}{\# $X$: Feature maps;}
        \STATE \textcolor{OliveGreen}{\# $X$.shape: ($C$, $H$, $W$); $C$: embedding channel; $H$: height; $W$: width;}
        \STATE \textcolor{OliveGreen}{\# $j$: index of STL (from $0$); $W_s$: size of window; $N_h$: number of heads.}
        \STATE \quad
        \STATE $N_w = HW / W_s^2$ $\quad$ \textcolor{OliveGreen}{\# Calculate the number of windows.}
        \STATE \quad
        \STATE $X_{\text{tmp}} \gets X$ $\quad$ \textcolor{OliveGreen}{\# For residual connection.}
        \STATE $X \gets \operatorname{LN}(X)$ $\quad$ \textcolor{OliveGreen}{\# Layer normalisation 1.}
        \STATE \quad
        \STATE \textcolor{OliveGreen}{\# Shifting operation used in even STL.}
        \IF{$j\%2 \neq 0$}
        \STATE $X \gets \operatorname{cyclic\_shift}(X)$
        \ENDIF 
        \STATE \textcolor{OliveGreen}{\# Split feature maps into non-overlapping windows.}
        \STATE $X_{\text{win}} \gets \operatorname{window\_partition}(X)$ $\quad$ \textcolor{OliveGreen}{\# $X_{\text{win}}$.shape: ($N_w N_h$, $C/N_h$, $W_s$, $W_s$)}
        \STATE \textcolor{OliveGreen}{\# Multi-head self-attention}
        \STATE $X_{\text{win}} \gets \operatorname{MSA}(X_{\text{win}})$ $\quad$ \textcolor{OliveGreen}{\# $X_{\text{win}}$.shape: ($N_w N_h$, $C/N_h$, $W_s$, $W_s$)}
        \STATE \textcolor{OliveGreen}{\# Recover feature maps from windows}
        \STATE $X \gets \operatorname{reverse\_window}(X_{\text{win}})$ $\quad$ \textcolor{OliveGreen}{\# $X$.shape: ($C$, $H$, $W$)}
        \STATE \textcolor{OliveGreen}{\# Corresponding shifting reversing operation used in even STL.}
        \IF{$j\%2 \neq 0$}
        \STATE $X \gets \operatorname{reverse\_shift}(X)$
        \ENDIF 
        \STATE $X \gets X + X_{\text{tmp}}$ $\quad$ \textcolor{OliveGreen}{\# Residual connection 1.}
        \STATE \quad
        \STATE $X_{\text{tmp}} \gets X$ $\quad$ \textcolor{OliveGreen}{\# For residual connection.}
        \STATE $X \gets \operatorname{LN}(X)$ $\quad$ \textcolor{OliveGreen}{\# Layer normalisation 2.}
        \STATE $X \gets \operatorname{MLP}(X)$ $\quad$ \textcolor{OliveGreen}{\# Multi-layer perceptron.}
        \STATE $X \gets X + X_{\text{tmp}}$ $\quad$ \textcolor{OliveGreen}{\# Residual connection 2.}
        \STATE \quad
    	\ENSURE $X$
	\end{algorithmic}}
\end{algorithm}

\begin{algorithm}[H]
	\renewcommand{\algorithmicrequire}{\textbf{Input:}}
	\renewcommand{\algorithmicensure}{\textbf{Output:}}
	\caption{(Shifted) windows multi-head self-attention.} 
	\label{alg:swmsa} 
	\setlength{\tabcolsep}{4mm}{
    \begin{algorithmic}
        \REQUIRE $X$
        \STATE \textcolor{OliveGreen}{\# $X$: windows for multi-head self-attention operation;}
        \STATE \textcolor{OliveGreen}{\# $X$.shape: ($N_w N_h$, $C/N_h$, $W_s$, $W_s$);}
        \STATE \textcolor{OliveGreen}{\# $N_w$: number of windows; $N_h$: number of heads;} 
        \STATE \textcolor{OliveGreen}{\# $C$: embedding channel; $W_s$: size of window.}
        \STATE \quad
        \STATE \textcolor{OliveGreen}{\# Calculate the query, key and value.}
    	\STATE $Q \gets \operatorname{Linear}_q(X)$
    	\STATE $K \gets \operatorname{Linear}_k(X)$
    	\STATE $V \gets \operatorname{Linear}_v(X)$
        \STATE \quad
        \STATE \textcolor{OliveGreen}{\# Calculate the relative position bias.}
    	\STATE $B \gets \operatorname{get\_relative\_position}(X)$
        \STATE \quad
        \STATE \textcolor{OliveGreen}{\# Calculate the attention result.}
    	\STATE $\text{attn\_map} \gets \operatorname{dot}(Q,K\text{.transpose})/\sqrt{C}$
    	\STATE $\text{attn\_map} \gets \operatorname{SoftMax}(\text{attn\_map}+B)$
        \STATE $\text{attn} \gets \operatorname{dot}(\text{attn\_map},V)$
        \STATE $\text{attn} \gets \operatorname{Linear}(\text{attn})$
        \STATE \quad
    	\ENSURE $\text{attn}$
	\end{algorithmic}}
\end{algorithm}

\subsubsection{Loss Function}
% Swin Transformer
A novel multi-channel loss using the sensitivity maps was introduced for better reconstruction quality and more textures and details.
Charbonnier loss~\citep{Lai2019} was utilised for the pixel-wise loss and the frequency loss since it is more robust and able to handle the outliers better.
The total loss $\mathcal{L}_{\mathrm{TOTAL}}(\theta)$ consists of the pixel-wise Charbonnier loss $\mathcal{L}_{\mathrm{pixel}}(\theta)$, the frequency Charbonnier loss $\mathcal{L}_{\mathrm{freq}}(\theta)$ and perceptual loss $\mathcal{L}_{\mathrm{VGG}}(\theta)$. The pixel-wise Charbonnier loss can be expressed by

\begin{align}\label{formula:14}
\mathop{\text{min}}\limits_{\theta} 
\mathcal{L}_{\mathrm{pixel}}(\theta) =
\frac{1}{S} \sum_{q=1}^{S} 
\sqrt{\mid\mid x^q - \mathcal{S}^q \hat x_u \mid\mid^2_2 + \epsilon^2},
\end{align}

\noindent where $\epsilon$ is a constant which is set to $10^{-9}$ empirically and $\mathcal{S}^q$ is the sensitivity map of $q^{\text{th}}$ coil ($S$ colis in total). The frequency Charbonnier loss can be expressed by

\begin{align}\label{formula:15}
\mathop{\text{min}}\limits_{\theta} 
\mathcal{L}_{\mathrm{freq}}(\theta) =
\frac{1}{S} \sum_{q=1}^{S} 
\sqrt{\mid\mid y^q - \mathcal{F}\mathcal{S}^q \hat x_u \mid\mid^2_2 + \epsilon^2}.
\end{align}

The perceptual VGG loss can be expressed by
\begin{align}\label{formula:16}
\mathop{\text{min}}\limits_{\theta} 
\mathcal{L}_{\mathrm{VGG}}(\theta) =
\mid\mid f_{\mathrm{VGG}}(x) - f_{\mathrm{VGG}}(\hat x_u) \mid\mid_1,
\end{align}

\noindent where $f_{\mathrm{VGG}}(\cdot)$ denotes the VGG network, and $\mid\mid \cdot \mid\mid_1$ denotes the $l_1$ norm. The utilisation of $\mathcal{L}_{\mathrm{VGG}}$ is able to optimise the perceptual quality of reconstructed results.

The total loss can be expressed by
\begin{align}\label{formula:17}
\mathcal{L}_{\mathrm{TOTAL}}(\theta)
= \alpha \mathcal{L}_{\mathrm{pixel}}(\theta)
+ \beta \mathcal{L}_{\mathrm{freq}}(\theta)
+ \gamma \mathcal{L}_{\mathrm{VGG}}(\theta),
\end{align}

\noindent where $\alpha$, $\beta$ and $\gamma$ are coefficients controlling the balance of each term in the loss function.

\section{Experiments and Results}

\subsection{Datasets}

\textcolor{black}{
In this work, the Calgary Campinas multi-channel (CC) dataset\footnote{https://sites.google.com/view/calgary-campinas-dataset/mr-reconstruction-challenge}~\citep{Souza2018} and the Multi-modal Brain Tumour Segmentation Challenge 2017 (BraTS17)\footnote{https://www.med.upenn.edu/sbia/brats2017/data.html}~\citep{BraTS17_1,BraTS17_2,BraTS17_3} dataset were used for the experiment sections.
}

\textcolor{black}{
The available data of the CC dataset contains 67 cases of three-dimensional (3D), 12-channel (117 scans), T1-weighted, gradient-recalled echo, 1 mm isotropic sagittal acquisitions. 
Acquisition parameters were TR/TE/TI = 6.3ms/2.6ms/650ms (93 scans) and TR/TE/TI = 7.4ms/3.1ms/400ms (74 scans), with 170 to 180 contiguous 1.0-mm slices and a field of view of 256mm $\times$ 218mm. 
The original CC dataset provides a hybrid $(x, k_y, k_z, C)$ structure ($x$: read-out direction; $y$: phase-encoding direction; $z$: slice-encoding direction; $C$: channels), where inverse Fourier transform is performed in the read-out direction. 
These 3D hybrid data were uniformly zero-filled in the phase-encoding direction to $256 \times 256$, and turned into 3D image space volumes by 2D iFFT on the $k_y-k_z$ plane. 
We randomly chose 40 cases for training, 7 cases for validation and 20 cases for testing, according to the ratio of 6:1:3 approximately. In each case, we chose 100 2D slices near the centre along the read-out direction (sagittal view).} 

% BraTS17
\textcolor{black}{
For the BraTS17 dataset, we applied the brain data with reference segmentation results (280 3D volumes in BraTS17 official training dataset), including both higher and lower grade glioma. 
These multi-modal scans contain native T1-weighted (T1), T1-contrast enhanced (T1CE), T2-weighted (T2), and T2 Fluid Attenuated Inversion Recovery (FLAIR) data.
These 280 3D brain data were divided into training, validation and testing set (235, 20, and 30 cases respectively), and cropped to $152 \times 192 \times 144$ volumes (slice, height and width, respectively). 
For each case, we used 100 slices near the centre in the training stage to avoid invalid data, i.e., slices that are totally dark or with little information, for training.}

\subsection{Implementation Detail}

The proposed SwinMR was implemented using PyTorch, trained on two NVIDIA RTX 3090 GPUs with 24GB GPU memory, and tested on an NVIDIA RTX 3090 GPU or an Intel Core i9-10980XE CPU. 
We set the RSTB number, the STL number, the window size number and the attention head number to 6, 6, 8 and 6 respectively, which are the default setting in the original SwinIR~\citep{Liang2021}. The patch number and channel number were empirically set to 96 and 180, according to our ablation studies. 
For the parameter in the loss function, $\alpha$, $\beta$, $\gamma$ were set to 15, 0.1 and 0.0025 to balance each term, according to our ablation studies. 
\textcolor{black}{
Our proposed SwinMR was trained for 100k steps using Adam optimiser.
The initial learning rate was set to $2 \times 10^{-4}$ and decayed by 0.5 every 10,000 steps from the 50,000$^{\text{th}}$ step. Random flip and rotation were applied for data augmentation.
}

We used SwinMR (PI) to denote the proposed model trained with multi-channel data and sensitivity maps, and SwinMR (nPI) to indicate the proposed model trained with single-channel data without sensitivity maps.

\subsection{Evaluation Methods}

Structural similarity index (SSIM), Peak signal-to-noise ratio (PSNR) and Fr\'echet inception distance (FID)~\citep{Heusel2017} were utilised for evaluation. 
SSIM quantifies the structural similarity between two images based on luminance, contrast, and structures.
PSNR is the ratio between maximum signal power and noise power, which measures the fidelity of the representation. Both metrics are based on simple and shallow functions, and direct comparisons between images, which are not necessary for the visual quality for human observers~\citep{Zhang2018}.
FID is calculated by computing the Fr\'echet distance between two multivariate Gaussians, which measures the similarity between two sets of images. FID correlates well with visual quality for human observers, and a lower FID indicates more perceptual results.

Both Intersection over Union (IoU) and Dice scores were applied to measure the segmentation quality in the brain tumour segmentation experiment. 

\textcolor{black}{The number of parameters (\#PARAMs) and Multiply-Accumulate Operations (MACs) were applied to measure the model size and the computational cost. MACs were calculated using a $1 \times 1 \times 256 \times 256$ array as input (Batch $\times$ Channel $\times$ Height $\times$ Width).}

\subsection{Comparisons with Other Methods}

In this experimental study, we compared our proposed SwinMR (nPI and PI) with other benchmarked MR reconstruction methods, including Deep ADMM Net~\citep{Yang2016}, \textcolor{black}{U-Net~\citep{Ronneberger2015}}, DAGAN~\citep{Yang2018}, PIDDGAN~\citep{Huang2021}, as well as ground truth MR images (GT) and undersampled zero-filled MR images (ZF) using Gaussian 1D 30\% mask. Among them, PIDDGAN and SwinMR (PI) were parallel imaging-coupled, i.e., trained with multi-channel MR images. This experiment was conducted using the CC dataset.

The quantitative result of comparisons is shown in Table~\ref{tab:comparison}. Our proposed SwinMR (nPI) achieved the highest SSIM and PSNR, and SwinMR (PI) achieved the best FID score. \textcolor{black}{The inference time} in Table~\ref{tab:comparison} indicates the average time for one inference measured by ten times inferences in average in an Intel Core i9-10980XE CPU or an NVIDIA RTX 3090 GPU. The computational cost of SwinMR was higher than other CNN-based models. \textcolor{black}{SwinMR has a larger computational cost (MACs) than other CNN-based and GAN-based methods, but with a smaller model size (\#PARAMs).} 

\begin{figure}[H]
    \centering
    \includegraphics[width=5in]{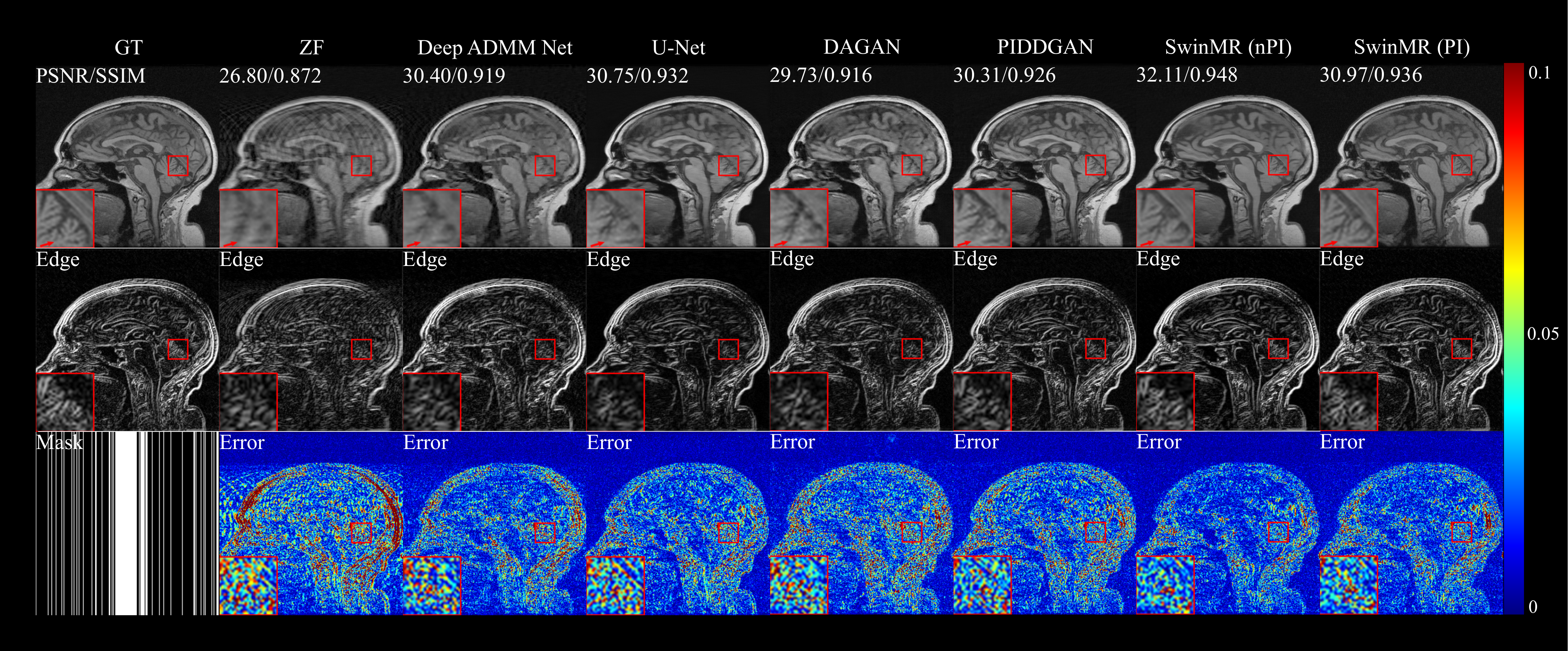}
    \caption{
    \textcolor{black}{
    Samples of the comparison experiment with ground truth images (GT), undersampled zero-filled images (ZF) and reconstructed images by other methods. 
    Row 1: GT, ZF and reconstructed images by different methods; 
    Row 2: Edge information extracted by Sobel operator; 
    Row 3: Gaussian 1D 30\% mask and the absolute differences between reconstructed (or ZF) images and GT images ($10 \times$).
    }}
    \label{fig:FIG_EXP_IMAGE_Comparison}
\end{figure}

Figure~\ref{fig:FIG_EXP_IMAGE_Comparison} shows the reconstructed MR images, edge information extracted by Sobel operator and absolute differences of standardised pixel intensities ($10 \times$) between reconstructed MR images and GT MR images from top to button respectively. The proposed SwinMR shows superiority to other methods in terms of overall reconstruction quality and edge information.

% Table comparison
\begin{table}[H]
    \centering
    \caption{\textcolor{black}{
    Quantitative results of the comparison experiment with other methods using Gaussian 1D 30\% mask (mean (std)). 
    $^{\dagger}$: $p < 0.05$;
    $^{\dagger\dagger}$: $p < 0.01$
    (compared with SwinMR (PI) by paired t-Test).
    $^{\ddagger}$: $p < 0.05$;
    $^{\ddagger\ddagger}$: $p < 0.01$
    (compared with SwinMR (nPI) by paired t-Test).
    $^{\star}$: \#PARAMs for only the generator/for both the generator and discriminator.
    PSNR: Peak signal-to-noise ratio;
    SSIM: Structural similarity index;
    FID: Fr\'echet inception distance;
    Inference Time: The average time for one inference in an Intel Core i9-10980XE CPU or an NVIDIA RTX 3090 GPU;
    \#PARAMs: The parameters number of models;
    MACs: Multiply-Accumulate Operations.\\
    }}
    % \resizebox{160mm}{!}{
    \scalebox{0.70}{
    \begin{tabular}{cccccccc}
        \toprule
        \multirow{2}[4]{*}{Methods} & \multirow{2}[4]{*}{PSNR} & \multirow{2}[4]{*}{SSIM} & \multirow{2}[4]{*}{FID} & \multicolumn{2}{c}{Inference Time} & \#PARAMs & MACs \\
    \cmidrule{5-6}          &       &       &       & CPU (s) & GPU (s) & (M)   & (G) \\
        \midrule
        ZF    & 27.81 (0.83)$^{\dagger\dagger\ddagger\ddagger}$ & 0.884 (0.012)$^{\dagger\dagger\ddagger\ddagger}$ & 156.39  & -     & -     & -     & - \\
        Deep ADMM Net & 29.24 (0.99)$^{\dagger\dagger\ddagger\ddagger}$ & 0.922 (0.012)$^{\dagger\dagger\ddagger\ddagger}$ & 54.56  & 0.459 (0.052) & -     & -     & - \\
        U-Net & 31.48 (0.86)$^{\dagger\dagger\ddagger\ddagger}$ & 0.939 (0.009)$^{\dagger\dagger\ddagger\ddagger}$ & 46.90  & 0.166 (0.007) & 0.006 (0.000) & 32.31  & 56.44  \\
        DAGAN & 30.41 (0.83)$^{\dagger\dagger\ddagger\ddagger}$ & 0.924 (0.010)$^{\dagger\dagger\ddagger\ddagger}$ & 56.05  & 0.089 (0.003) & 0.003 (0.000) & 98.59/127.18$^{\star}$ & 33.97  \\
        PIDDGAN & 31.23 (0.93)$^{\dagger\dagger\ddagger\ddagger}$ & 0.936 (0.010)$^{\dagger\dagger\ddagger\ddagger}$ & 17.55  & 0.166 (0.007) & 0.006 (0.000) & 32.31/89.50$^{\star}$ & 56.44  \\
        \midrule
        SwinMR (nPI) & \textbf{33.06 (1.09)$^{\dagger\dagger}$} & \textbf{0.956 (0.009)$^{\dagger\dagger}$} & 21.03  & 19.310 (0.115) & 0.041 (0.001) & 11.40  & 800.73  \\
        SwinMR (PI) & 32.07 (1.02)$^{\ddagger\ddagger}$ & 0.945 (0.010)$^{\ddagger\ddagger}$ & \textbf{8.70} & 19.310 (0.115) & 0.041 (0.001) & 11.40  & 800.73  \\
        \bottomrule
    \end{tabular}}%
    \label{tab:comparison}%
\end{table}%

\subsection{Experiments on Masks}

This experimental study aimed to evaluate the performance of SwinMR using different undersampling trajectories. 
Three 1D Cartesian undersampling trajectories including Gaussian 1D 10\% (G1D10\%), Gaussian 1D 30\% (G1D30\%) and Gaussian 1D 50\% (G1D50\%), as well as two 2D non-Cartesian undersampling trajectories including radial 10\% (R10\%) and spiral 10\% (S10\%) were applied in this experiment. This experiment compared the SSIM, PSNR and FID of SwinMR (PI), DAGAN and ZF, and was conducted using the CC dataset.

\begin{figure}[H]
    \centering
    \includegraphics[width=5in]{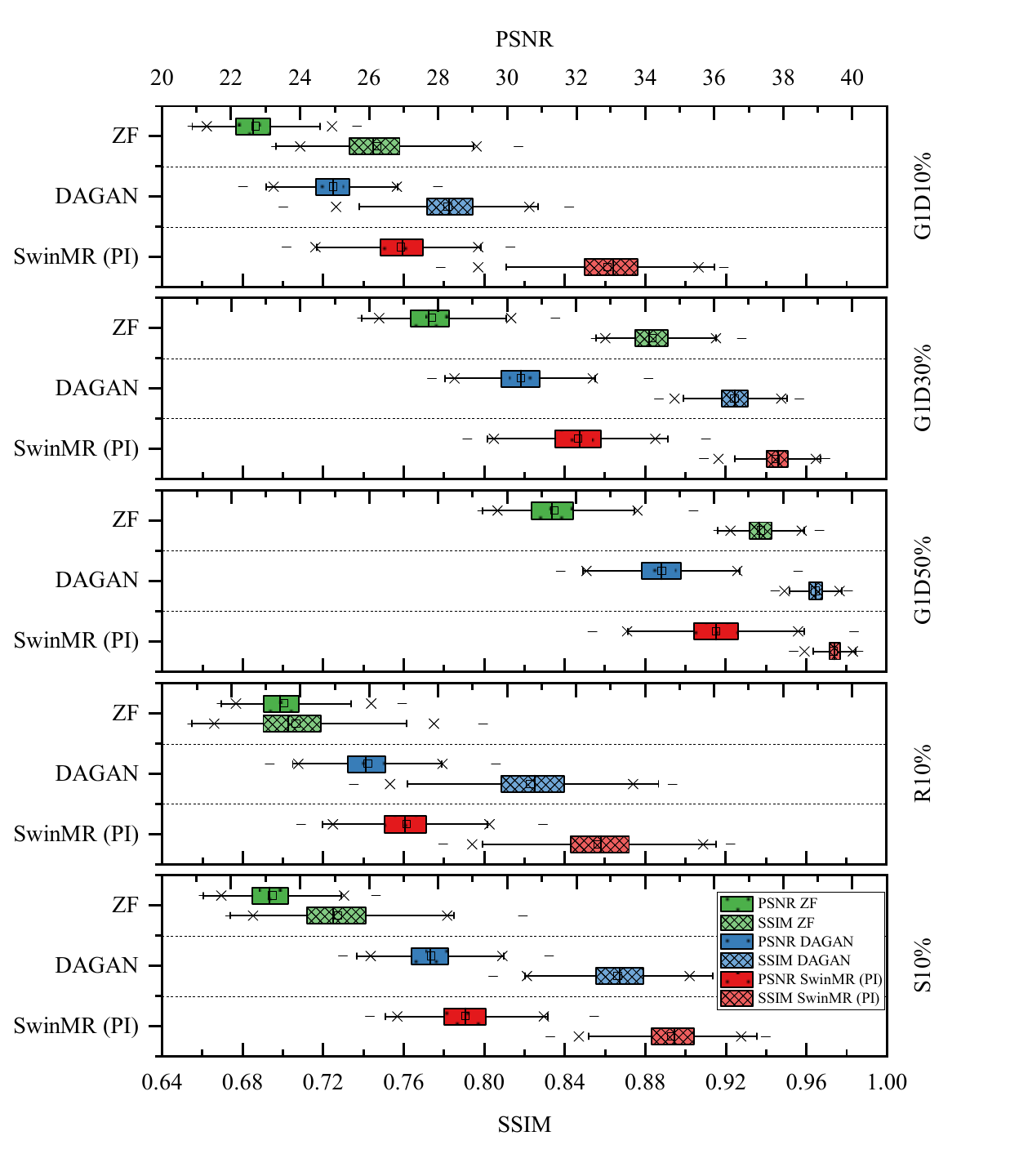}
    \caption{
    Peak signal-to-noise ratio (PSNR) and Structural similarity index (SSIM) of the experiment on different masks. 
    Five undersampling trajectories including Gaussian 1D 10\% (G1D10\%), Gaussian 1D 30\% (G1D30\%), Gaussian 1D 50\% (G1D50\%), radial 10\% (R10\%) and spiral 10\% (S10\%) were applied in this experiment. 
    (Box range: interquartile range; $\times$:1\% and 99\% confidence interval; $-$: maximum and minimum; $\square$: mean; $\shortmid$: median.) The SwinMR (PI) outperforms the DAGAN using different undersampling masks with significantly higher PSNR, SSIM ($p < 0.05$ by paired t-Test).}
    \label{fig:FIG_EXP_GRAPH_Mask}
\end{figure}

\begin{figure}[H]
    \centering
    \includegraphics[width=5in]{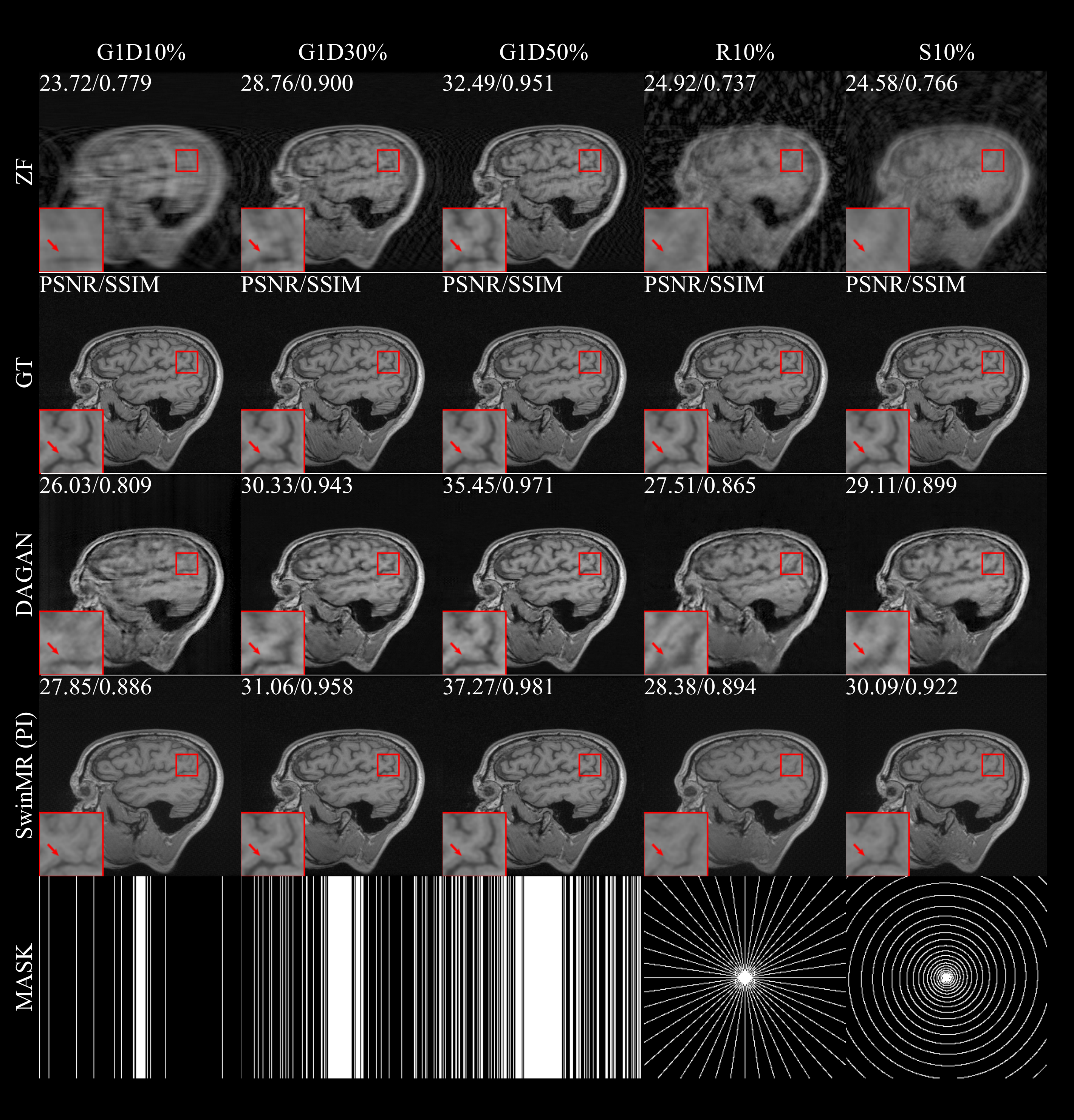}
    \caption{
    Samples of the experiment on different masks. 
    Five undersampling trajectories including Gaussian 1D 10\% (G1D10\%), Gaussian 1D 30\% (G1D30\%), Gaussian 1D 50\% (G1D50\%), radial 10\% (R10\%) and spiral 10\% (S10\%) were applied in this experiment.
    Row 1: Undersampled zero-filled MR images (ZF) using different masks;
    Row 2: Ground truth MR images (GT);
    Row 3: Reconstructed MR images by DAGAN;
    Row 4: Reconstructed MR images by SwinMR (PI);
    Row 5: Undersampling masks.
    The Peak signal-to-noise ratio (PSNR) and Structural similarity index (SSIM) of reconstructed and ZF images are shown in the top-left corner.
    }
    \label{fig:FIG_EXP_IMAGE_Mask}
\end{figure}

\begin{figure}[H]
    \centering
    \includegraphics[width=5in]{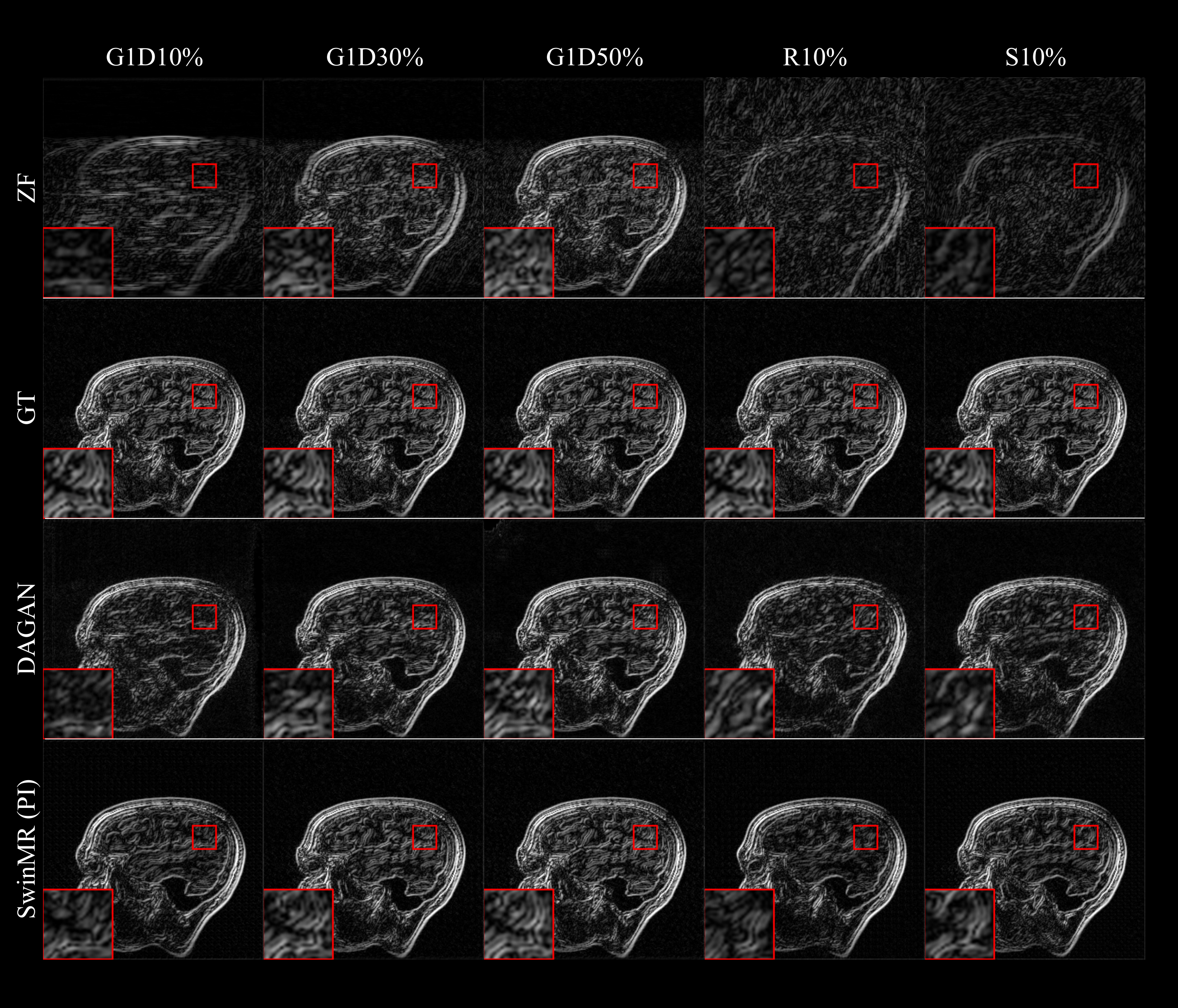}
    \caption{
    Edge information of the experiment on different masks. 
    Five undersampling trajectories including Gaussian 1D 10\% (G1D10\%), Gaussian 1D 30\% (G1D30\%), Gaussian 1D 50\% (G1D50\%), radial 10\% (R10\%) and spiral 10\% (S10\%) were applied in this experiment.
    Row 1: Edge information of undersampled zero-filled MR images (ZF) using different masks;
    Row 2: Edge information of ground truth MR images (GT);
    Row 3: Edge information of reconstructed MR images by DAGAN;
    Row 4: Edge information of reconstructed MR images by SwinMR (PI).
    The edge information was extracted by the Sobel operator.
    }
    \label{fig:FIG_EXP_EDGE_Mask}
\end{figure}

\begin{figure}[H]
    \centering
    \includegraphics[width=5in]{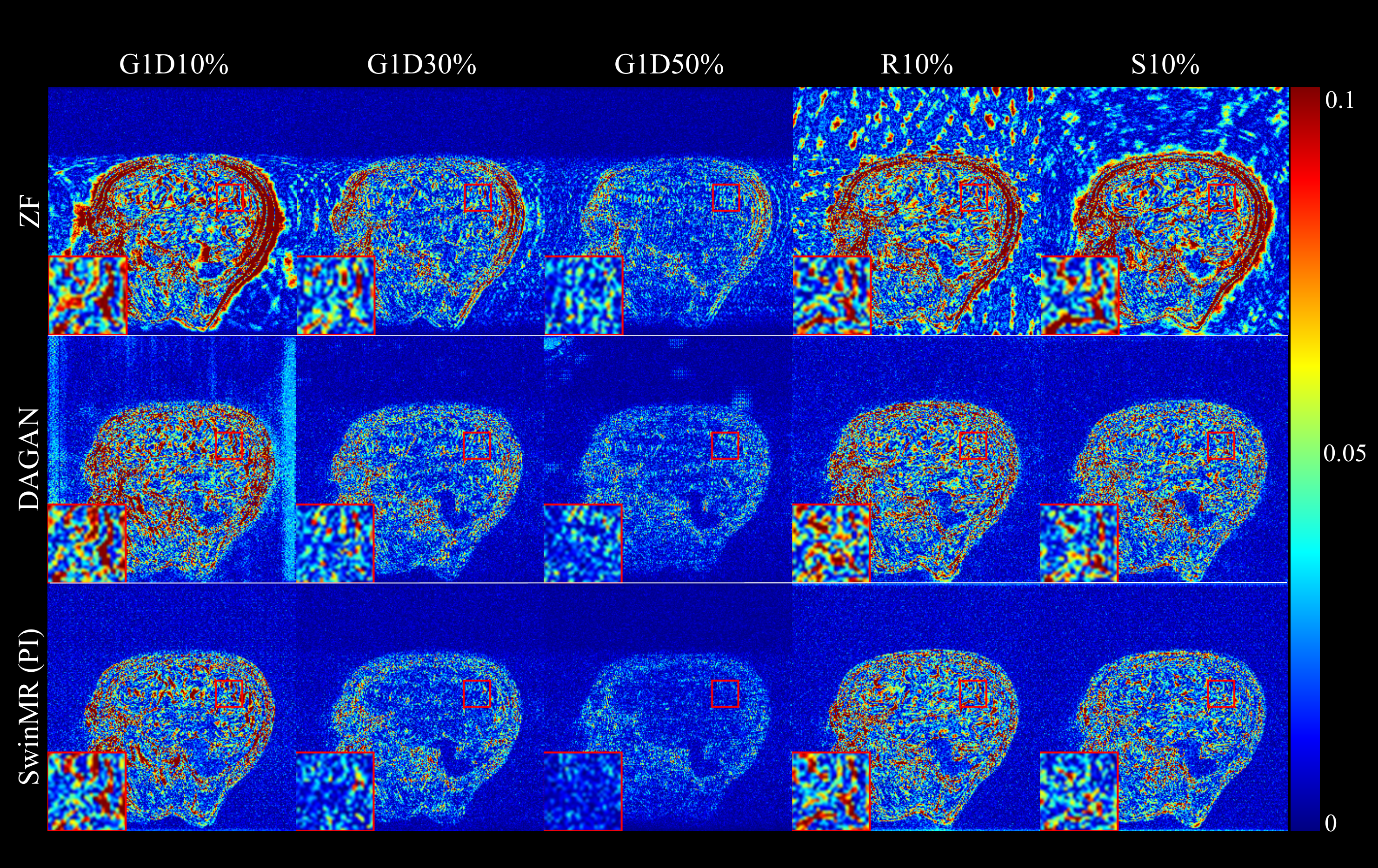}
    \caption{
    Absolute differences of standardised pixel intensities ($10 \times$) of the experiment on different masks. 
    Five undersampling trajectories including Gaussian 1D 10\% (G1D10\%), Gaussian 1D 30\% (G1D30\%), Gaussian 1D 50\% (G1D50\%), radial 10\% (R10\%) and spiral 10\% (S10\%) were applied in this experiment.
    Row 1: Absolute differences between undersampled zero-filled MR images (ZF) using different masks and ground truth MR images (GT);
    Row 2: Absolute differences between reconstructed MR images by DAGAN and GT;
    Row 3: Absolute differences between reconstructed MR images by SwinMR (PI) and GT.
    }
    \label{fig:FIG_EXP_ERROR_Mask}
\end{figure}

The quantitative results of the experiment on masks are shown in Figure~\ref{fig:FIG_EXP_GRAPH_Mask} and Table~\ref{tab:mask_fid}.
The sample of reconstructed images, edge information and absolute differences of standardised pixel intensities ($10 \times$) between reconstructed images and GT images are shown in Figure~\ref{fig:FIG_EXP_IMAGE_Mask}, Figure~\ref{fig:FIG_EXP_EDGE_Mask} and Figure~\ref{fig:FIG_EXP_ERROR_Mask} respectively. 
According to the results, the proposed SwinMR achieved a higher reconstruction quality compared to DAGAN using different undersampling trajectories, especially when the mask of low undersampling rate (10\%) was applied. 

% Mask FID
\begin{table}[H]
    \centering
    \caption{
    Fr\'echet inception distance (FID) of the experiment on different masks.
    Five undersampling masks including Gaussian 1D 10\% (G1D10\%), Gaussian 1D 30\% (G1D30\%), Gaussian 1D 50\% (G1D50\%), radial 10\% (R10\%) and spiral 10\% (S10\%) were applied in this experiment.\\
    }
    \setlength{\tabcolsep}{8mm}{
    \begin{tabular}{cccc}
        \toprule
        Mask  & SwinMR (PI) & DAGAN & ZF \\
        \midrule
        G1D10\% & \textbf{28.27 } & 169.83  & 326.00  \\
        G1D30\% & \textbf{8.70 } & 56.05  & 156.38  \\
        G1D50\% & \textbf{5.11 } & 19.26  & 86.25  \\
        R10\% & \textbf{34.19 } & 132.58  & 319.45  \\
        S10\% & \textbf{28.97 } & 115.98  & 333.40  \\
        \bottomrule
    \end{tabular}}%
    \label{tab:mask_fid}%
\end{table}%

\subsection{Experiments on Noise}

This experimental study aimed to evaluate the robustness of SwinMR under the influence of noise. 
The noise in MRI is imposed on the \textit{k}-space and that could follow a Gaussian distribution~\citep{Hansen2015}. In our experiments, different noise levels (NL20\%, NL30\%, NL50\%, NL70\% and NL80\%) were tested after undersampling (Gaussian 1D 30\% mask) in \textit{k}-space. The noise level is defined as:

\begin{align}\label{formula:18}
\text{NL} = \frac{N^{\prime}}{S^{\prime}+N^{\prime}},
\end{align}

\noindent where $N^{\prime}$ and $S^{\prime}$ denote the power of noise and signal, respectively. This experiment compared the SSIM, PSNR and FID of SwinMR (PI), DAGAN and ZF, and was conducted using the CC dataset.

\begin{figure}[H]
    \centering
    \includegraphics[width=5in]{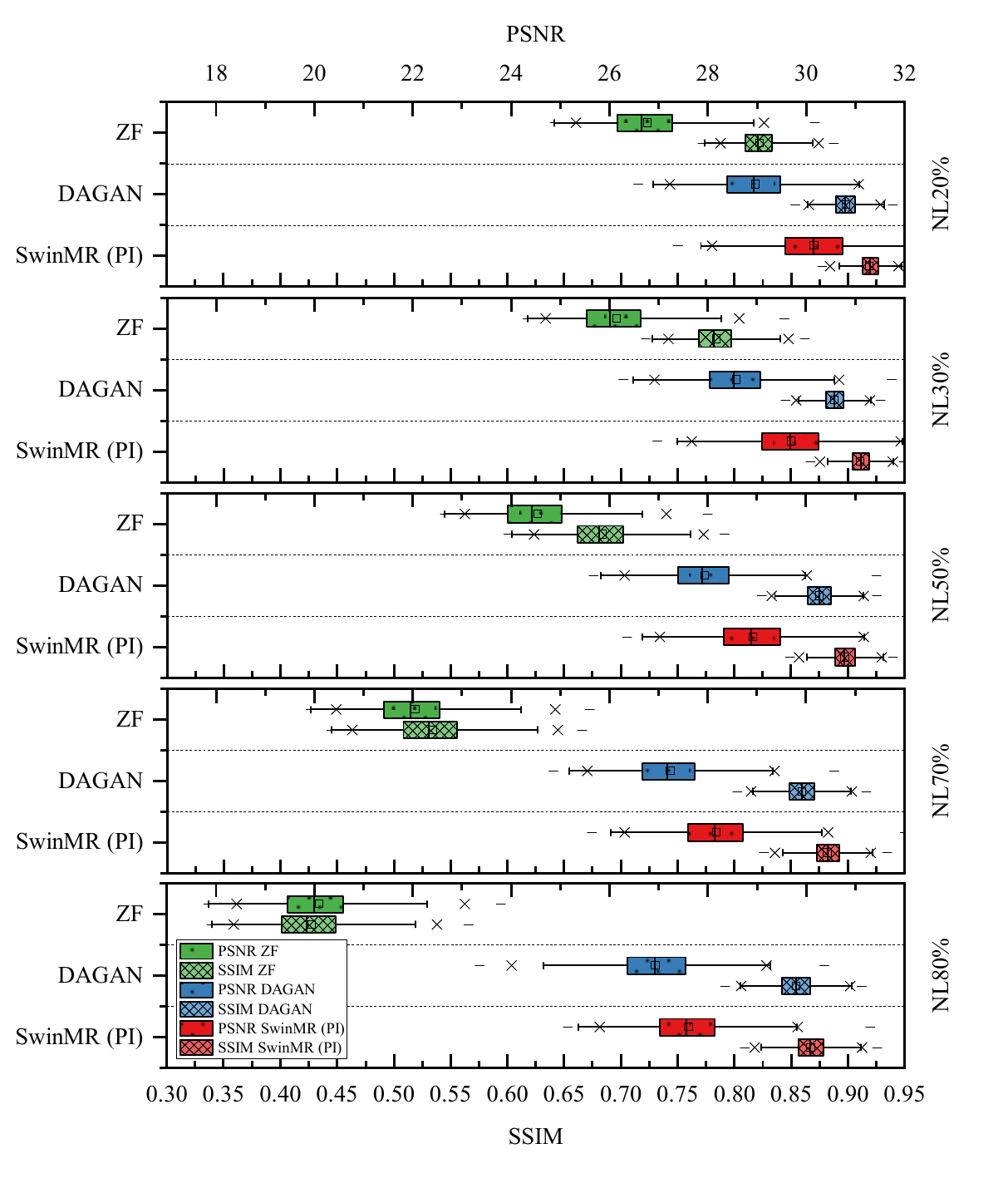}
    \caption{
    Peak signal-to-noise ratio (PSNR) and Structural similarity index (SSIM) of the experiment on different noise using Gaussian 1D 30\% mask. 
    Five noise levels (NL20\%, NL30\%, NL50\%, NL70\% and NL80\%) were tested in this experiment.
    (Box range: interquartile range; $\times$:1\% and 99\% confidence interval; $-$: maximum and minimum; $\square$: mean; $\shortmid$: median.)
    The SwinMR (PI) outperforms the DAGAN under different noise levels with significantly higher PSNR, SSIM ($p < 0.05$ by paired t-Test). 
    }
    \label{fig:FIG_EXP_GRAPH_Noise}
\end{figure}

\begin{figure}[H]
    \centering
    \includegraphics[width=5in]{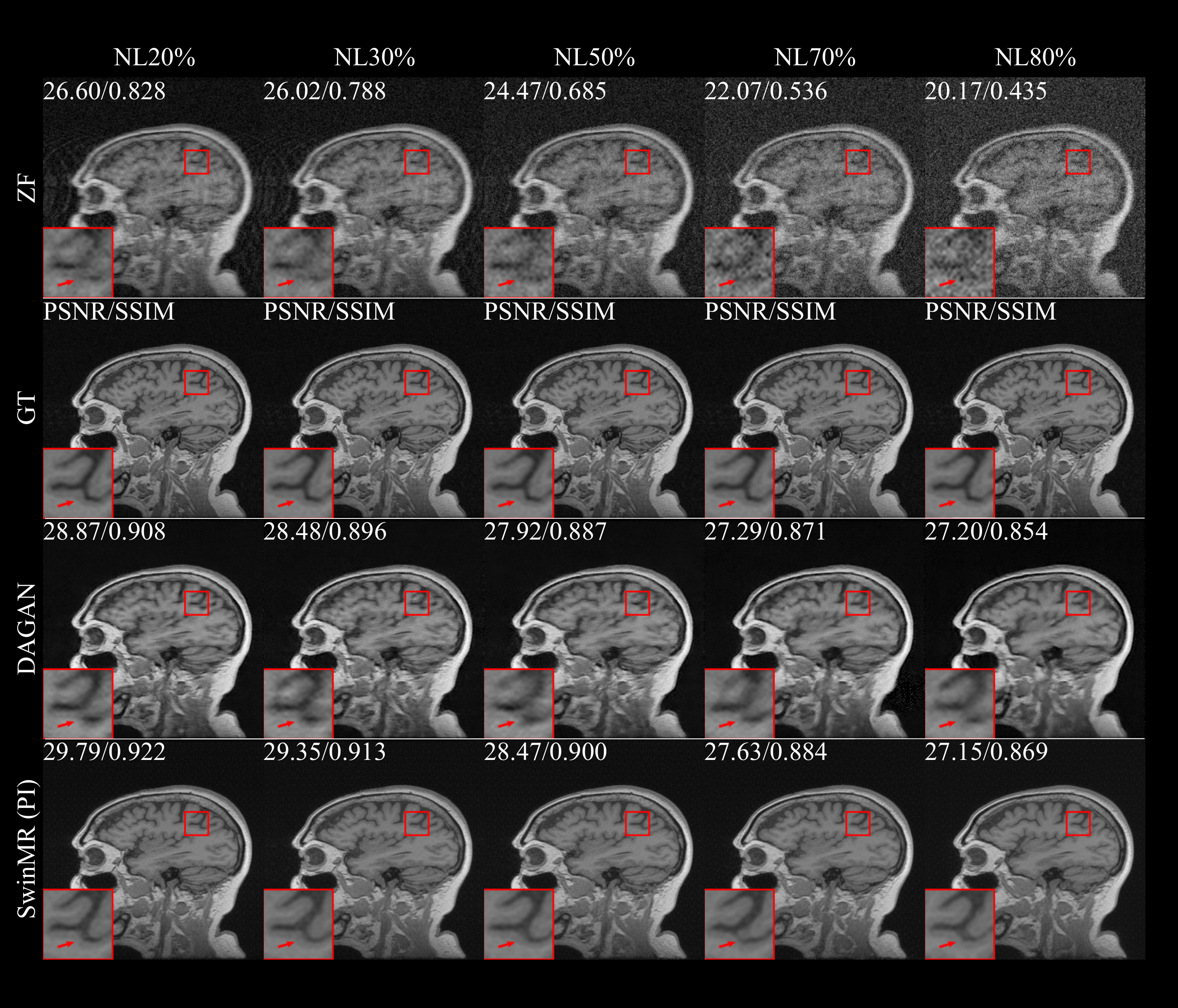}
    \caption{
    Samples of the experiment on different noise using Gaussian 1D 30\% mask.
    Five noise levels (NL20\%, NL30\%, NL50\%, NL70\% and NL80\%) were tested in this experiment.
    Row 1: Undersampled zero-filled MR images (ZF) with different noise levels;
    Row 2: Ground truth MR images (GT);
    Row 3: Reconstructed MR images by DAGAN;
    Row 4: Reconstructed MR images by SwinMR (PI).
    The Peak signal-to-noise ratio (PSNR) and Structural similarity index (SSIM) of reconstructed and ZF images are shown in the top-left corner.
    }
    \label{fig:FIG_EXP_IMAGE_Noise}
    \end{figure}

\begin{figure}[H]
    \centering
    \includegraphics[width=5in]{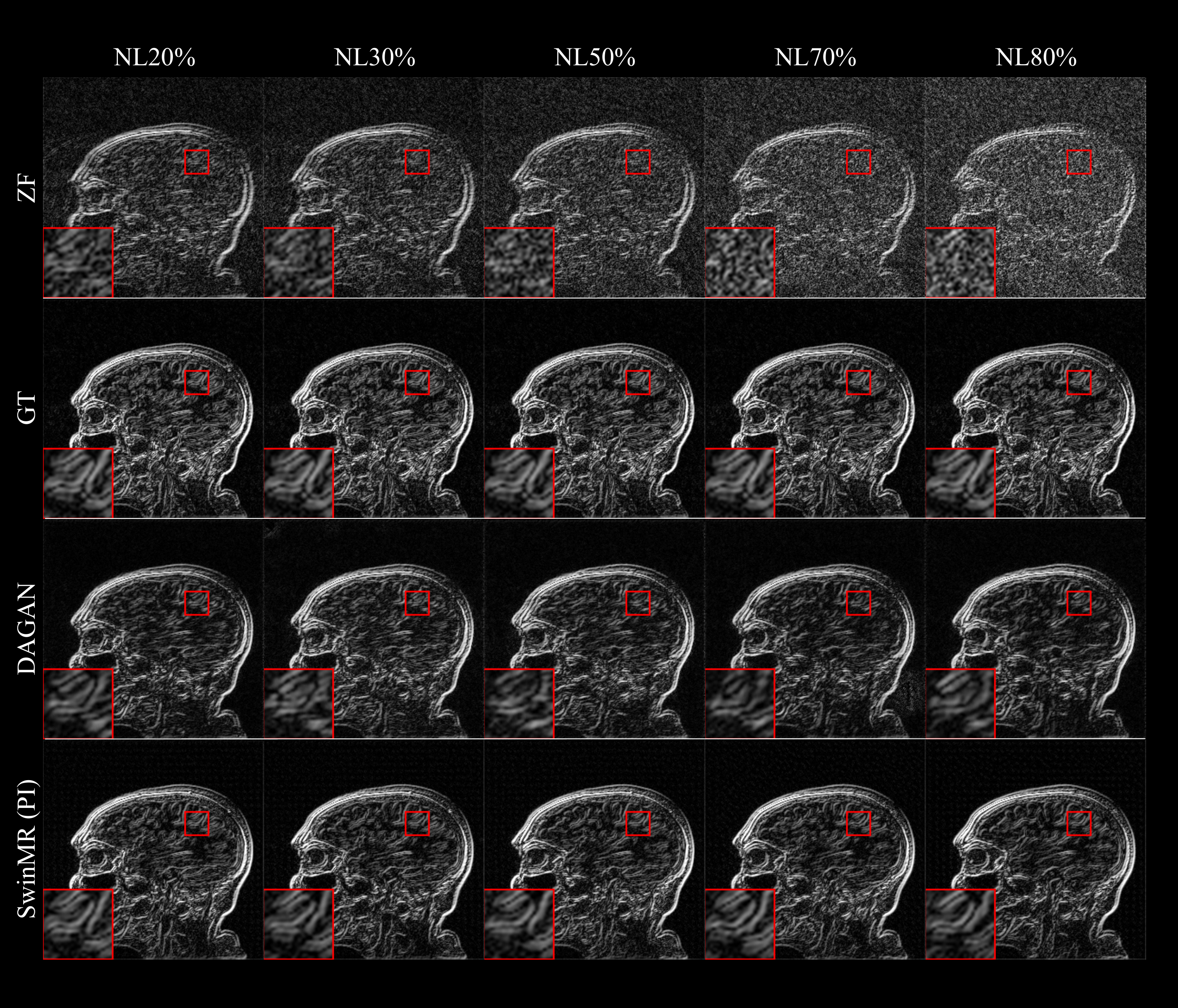}
    \caption{
    Edge information of the experiment on different noise using Gaussian 1D 30\% mask.
    Five noise levels (NL20\%, NL30\%, NL50\%, NL70\% and NL80\%) were tested in this experiment.
    Row 1: Edge information of undersampled zero-filled MR images (ZF) with different noise levels;
    Row 2: Edge information of ground truth images (GT);
    Row 3: Edge information of reconstructed MR images by DAGAN;
    Row 4: Edge information of reconstructed MR images by SwinMR (PI).
    The edge information was extracted by the Sobel operator.
    }
    \label{fig:FIG_EXP_EDGE_Noise}
\end{figure}

\begin{figure}[H]
    \centering
    \includegraphics[width=5in]{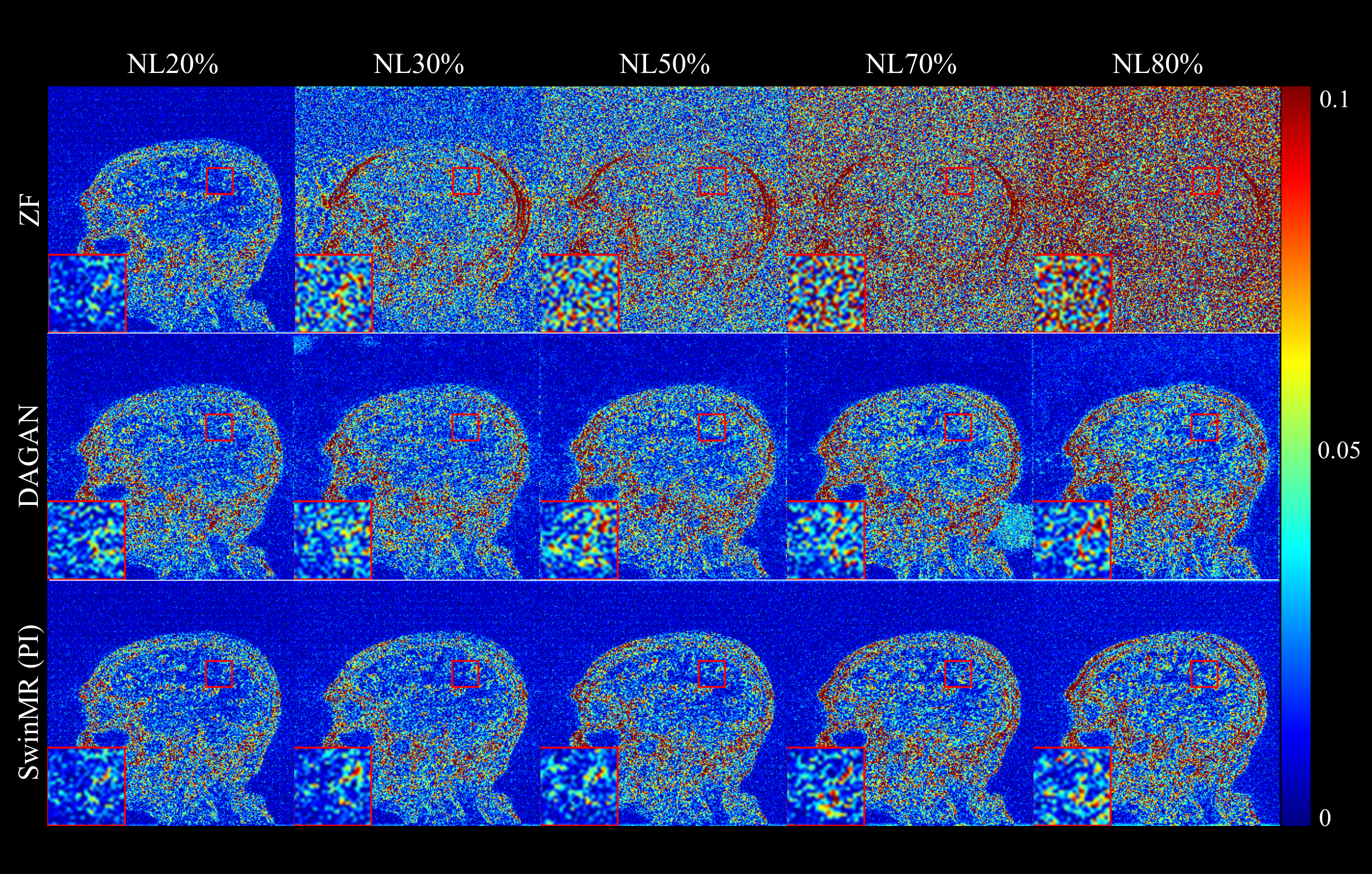}
    \caption{
    Absolute differences of standardised pixel intensities ($10 \times$) of the experiment on different using Gaussian 1D 30\% mask noise.
    Five noise levels (NL20\%, NL30\%, NL50\%, NL70\% and NL80\%) were tested in this experiment.
    Row 1: Absolute differences between undersampled zero-filled MR images (ZF) with different noise levels and ground truth MR images (GT);
    Row 2: Absolute differences between reconstructed MR images by DAGAN and GT;
    Row 3: Absolute differences between reconstructed MR images by SwinMR (PI) and GT.
    }
    \label{fig:FIG_EXP_ERROR_Noise}
\end{figure}

The quantitative results of the noise experiments are shown in Figure~\ref{fig:FIG_EXP_GRAPH_Noise} and Table~\ref{tab:noise_fid}.
The sample of reconstructed images, edge information and absolute differences of standardised pixel intensities ($10 \times$) between reconstructed images and GT images are shown in Figure~\ref{fig:FIG_EXP_IMAGE_Noise}, Figure~\ref{fig:FIG_EXP_EDGE_Noise} and Figure~\ref{fig:FIG_EXP_ERROR_Noise}, respectively.

According to the results, under the interruption of noise, SwinMR maintains better reconstruction quality compared to DAGAN.
The quality improvement becomes more clear when under a high noise level.

% Noise FID
\begin{table}[H]
    \centering
    \caption{
    Fr\'echet inception distance (FID) of the experiment on different noise using Gaussian 1D 30\% mask.
    Five noise levels (NL20\%, NL30\%, NL50\%, NL70\% and NL80\%) were applied in this experiment.\\
    }
    \setlength{\tabcolsep}{8mm}{
    \begin{tabular}{cccc}
        \toprule
        Noise Level & SwinMR (PI) & DAGAN & ZF \\
        \midrule
        NL20\% & \textbf{16.07 } & 66.60  & 156.77  \\
        NL30\% & \textbf{16.44 } & 71.50  & 168.61  \\
        NL50\% & \textbf{24.29 } & 75.32  & 203.39  \\
        NL70\% & \textbf{30.65 } & 85.32  & 251.15  \\
        NL80\% & \textbf{33.79 } & 80.97  & 282.40  \\
        \bottomrule
    \end{tabular}}%
    \label{tab:noise_fid}%
\end{table}%

\subsection{Ablation Experiments on the Patch Number and Channel Number}

The patch number $H$ (or $W$) and the channel number $C$ decide the input size of STL in the SwinMR. Ablation studies for different patch numbers and channel numbers were conducted to study the impression of them on the reconstruction results.

\begin{figure}[H]
    \centering
    \includegraphics[width=5in]{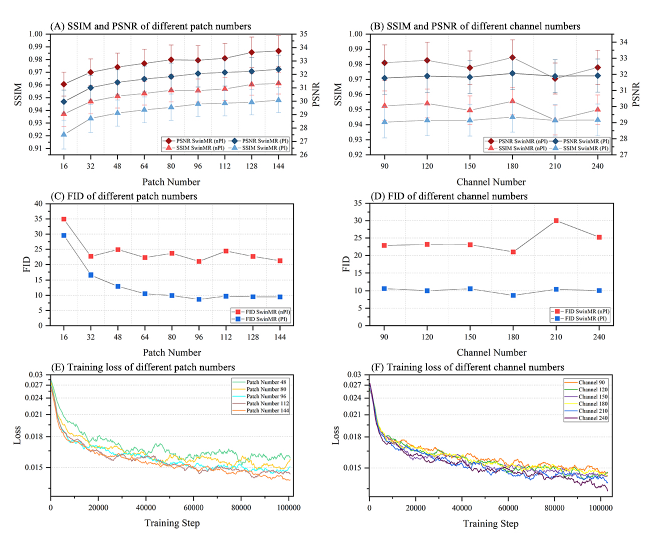}
    \caption{
    Structural similarity index (SSIM), Peak signal-to-noise ratio (PSNR), Fr\'echet inception distance (FID) \textcolor{black}{and training loss} of ablation experiments of the patch number and channel number. 
    (A), (C) and (E) are the SSIM/PSNR, FID and training loss of the ablation experiment of the patch number. 
    (B), (D) and (F) are the SSIM/PSNR, FID and training loss of the ablation experiment of the channel number.
    }
    \label{fig:FIG_AEXP_GRAPH_ChannelPatch}
\end{figure}

\begin{figure}[H]
    \centering
    \includegraphics[width=5in]{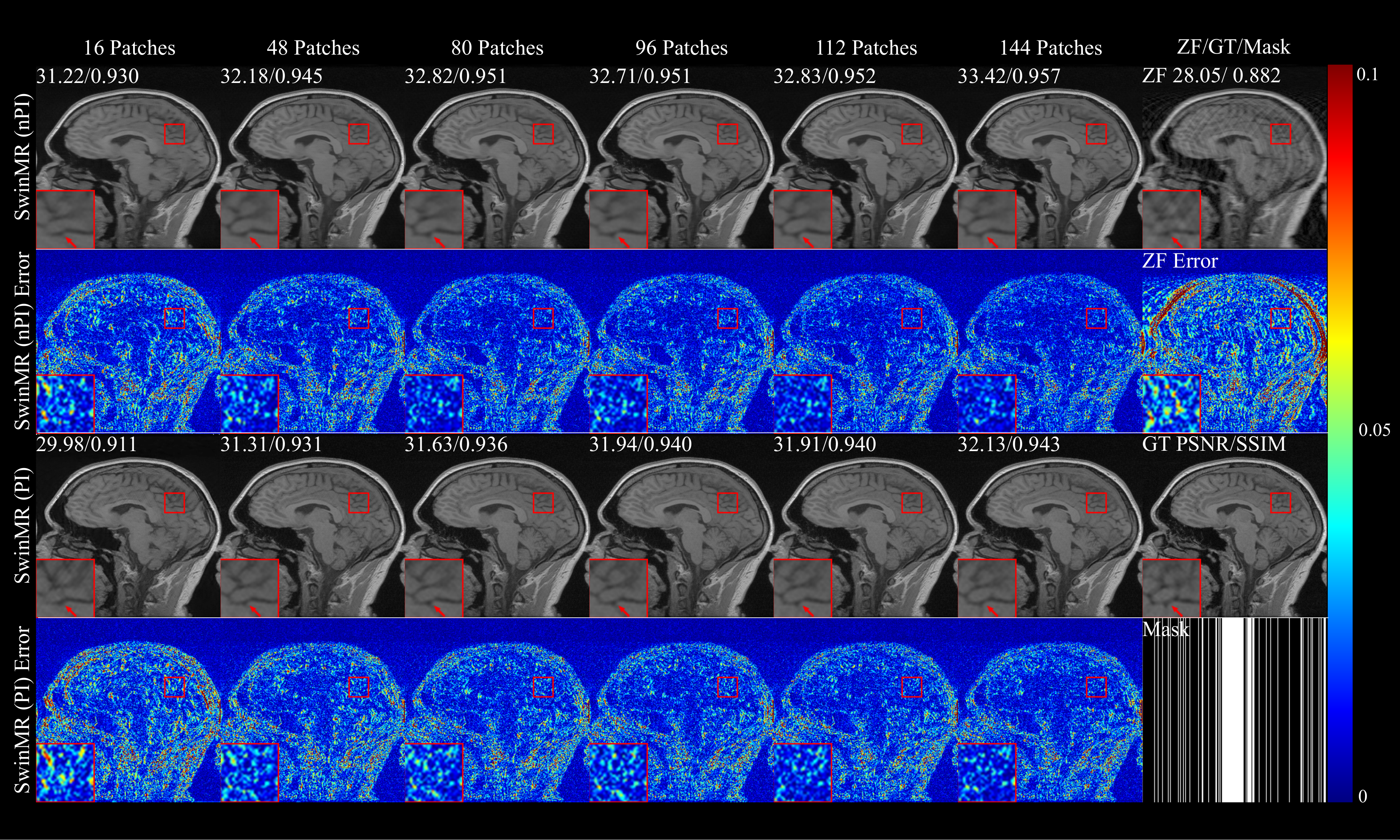}
    \caption{
    Samples of the ablation experiment on the patch number using Gaussian 1D 30\% mask.
    Row 1: Reconstructed MR images by SwinMR (nPI) with different patch numbers 
    and zero-filled MR images (ZF);
    Row 2: Absolute differences ($10 \times$) between reconstructed MR images by SwinMR (nPI) and ground truth MR images (GT),
    and absolute differences ($10 \times$) between ZF and GT;
    Row 3: Reconstructed MR images by SwinMR (PI) with the different patch number 
    and GT;
    Row 4: Absolute differences ($10 \times$) between reconstructed MR images by SwinMR (PI) and GT,
    and the Gaussian 1D 30\% mask.
    }
    \label{fig:FIG_AEXP_Patch}
\end{figure}

Figure~\ref{fig:FIG_AEXP_GRAPH_ChannelPatch} (A) and Figure~\ref{fig:FIG_AEXP_GRAPH_ChannelPatch} (C) show the SSIM, PSNR and FID of SwinMR with different patch numbers. 
Figure~\ref{fig:FIG_AEXP_GRAPH_ChannelPatch} (E) shows the loss function of SwinMR in the training process.
Figure~\ref{fig:FIG_AEXP_Patch} displays the sample of reconstructed images of SwinMR with different patch numbers. 

\begin{figure}[H]
    \centering
    \includegraphics[width=5in]{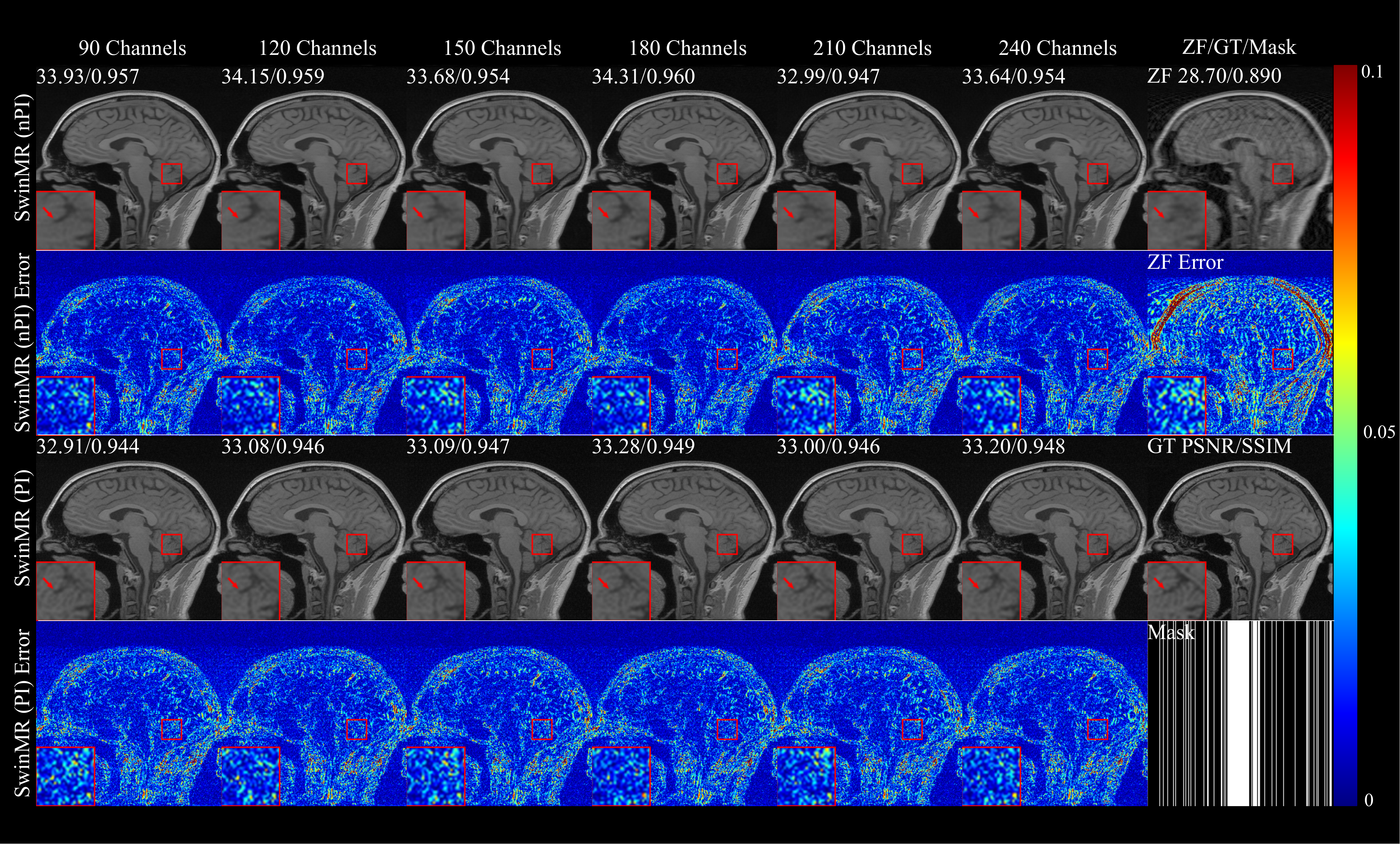}
    \caption{
    Samples of the ablation experiment on the channel number using Gaussian 1D 30\% mask.
    Row 1: Reconstructed MR images by SwinMR (nPI) with the different channel numbers 
    and zero-filled MR images (ZF);
    Row 2: Absolute differences ($10 \times$) between reconstructed MR images by SwinMR (nPI) and ground truth MR images (GT),
    and absolute differences ($10 \times$) between ZF and GT;
    Row 3: Reconstructed MR images by SwinMR (PI) with the different channel number 
    and GT;
    Row 4: Absolute differences ($10 \times$) between reconstructed MR images by SwinMR (PI) and GT,
    and the Gaussian 1D 30\% mask.
    }
    \label{fig:FIG_AEXP_Channel}
\end{figure}

Figure~\ref{fig:FIG_AEXP_GRAPH_ChannelPatch} (B) and Figure~\ref{fig:FIG_AEXP_GRAPH_ChannelPatch} (D) show the SSIM, PSNR and FID of SwinMR with different channel numbers. 
Figure~\ref{fig:FIG_AEXP_GRAPH_ChannelPatch} (F) shows the loss function of SwinMR in the training process.
Figure~\ref{fig:FIG_AEXP_Channel} displays the sample of reconstructed images of SwinMR with the different channel numbers. 

For the patch number, from Figure~\ref{fig:FIG_AEXP_GRAPH_ChannelPatch} (A) and Figure~\ref{fig:FIG_AEXP_GRAPH_ChannelPatch} (C), the results demonstrate that reconstruction quality becomes better as the patch number grows. According to Figure~\ref{fig:FIG_AEXP_GRAPH_ChannelPatch} (E), the training loss converges faster and lower as the patch number grows. However, the growing patch number aggravates the computational cost. Empirically, we applied patch number 96 for training.

For the channel number, from Figure~\ref{fig:FIG_AEXP_GRAPH_ChannelPatch} (B) and Figure~\ref{fig:FIG_AEXP_GRAPH_ChannelPatch} (D), the results did not resemble the trend presented in the ablation experiment on patch number. There were no significant differences for the three \textcolor{black}{metrics} (SSIM, PSNR and FID) as the channel number changed. According to Figure~\ref{fig:FIG_AEXP_GRAPH_ChannelPatch} (F), the training loss converges faster and lower as the channel number grows. Empirically, we applied a \textcolor{black}{channel} number of 180 for training.

For the comparison of multi-channel data (PI) and single-channel data (nPI), SwinMR (PI) tend to have a better (lower) FID, but worse (lower) SSIM/PSNR than SwinMR (nPI).

\subsection{Ablation Experiments on the Loss Function}

This ablation study aimed to discover the effect of each term in the loss function. According to Equation (\ref{formula:17}), the loss function of SwinMR consists of pixel-wise loss, frequency loss and perceptual loss. Four experiments were performed in this ablation study: 
(1) PFP: \textbf{P}ixel-wise, \textbf{F}requency and \textbf{P}erceptual loss; 
(2) PP: \textbf{P}ixel-wise and \textbf{P}erceptual loss; 
(3) PF: \textbf{P}ixel-wise and \textbf{F}requency loss; 
(4) P: only \textbf{P}ixel-wise loss.

\begin{figure}[H]
    \centering
    \includegraphics[width=5in]{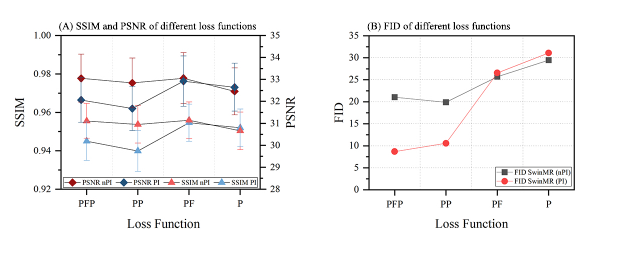}
    \caption{
    Structural similarity index (SSIM), Peak signal-to-noise ratio (PSNR) and Fr\'echet inception distance (FID) of the ablation experiment on the loss function using Gaussian 1D 30\% mask. 
    PFP: pixel-wise, frequency and perceptual loss; 
    PP: pixel-wise and perceptual loss; 
    PF: pixel-wise and frequency loss; 
    P: only pixel-wise loss.
    }
    \label{fig:FIG_AEXP_GRAPH_Loss}
\end{figure}

\begin{figure}[H]
    \centering
    \includegraphics[width=5in]{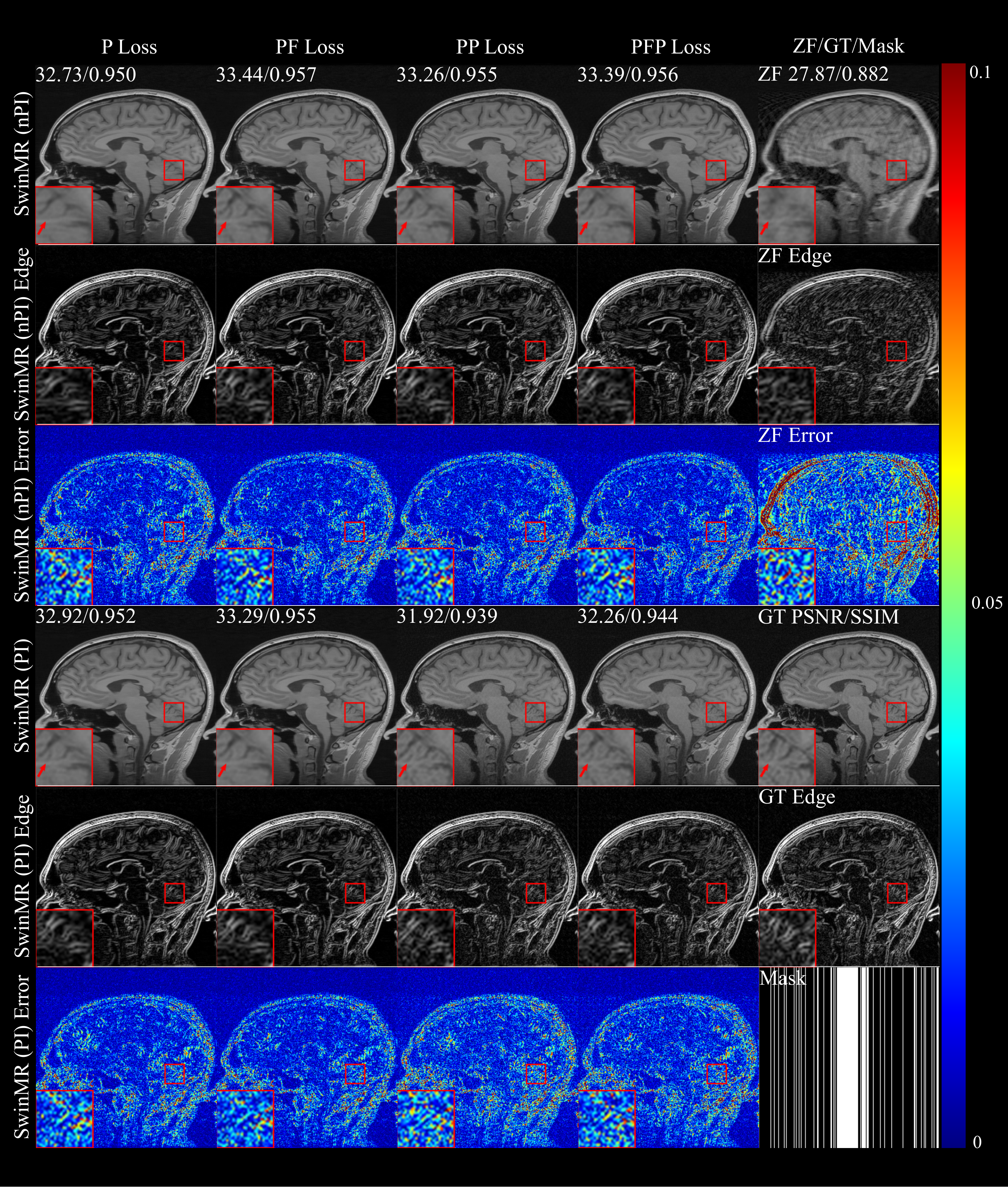}
    \caption{
    Samples of the ablation experiment on the loss function using Gaussian 1D 30\% mask.
    PFP: pixel-wise, frequency and perceptual loss; 
    PP: pixel-wise and perceptual loss; 
    PF: pixel-wise and frequency loss; 
    P: only pixel-wise loss.
    Row 1: Reconstructed MR images by SwinMR (nPI) 
    and zero-filled MR images (ZF);
    Row 2: Edge information of reconstructed MR images by SwinMR (nPI)
    and edge information of ZF; 
    Row 3: Absolute differences ($10 \times$) between reconstructed MR images by SwinMR (nPI) and ground truth MR images (GT),
    and absolute differences ($10 \times$) between ZF and GT;
    Row 4: Reconstructed MR images by SwinMR (PI)
    and GT;
    Row 5: Edge information of reconstructed MR images by SwinMR (PI)
    and edge information of GT; 
    Row 6: Absolute differences ($10 \times$) between reconstructed MR images by SwinMR (PI) and GT, 
    and the Gaussian 1D 30\% mask.
    }
    \label{fig:FIG_AEXP_Loss_edge}
\end{figure}

Figure~\ref{fig:FIG_AEXP_GRAPH_Loss} shows the SSIM, PSNR and FID of SwinMR trained with different loss functions.
Figure~\ref{fig:FIG_AEXP_Loss_edge} displays the samples of reconstructed images of SwinMR trained with different loss functions.

According to Figure~\ref{fig:FIG_AEXP_GRAPH_Loss}, for SwinMR (PI), the utilisation of frequency loss tends to improve SSIM/PSNR and decreases the FID (PFP vs PP; PF vs P). 
For SwinMR (nPI), the utilisation of frequency loss leads to improvement only on SSIM and PSNR, but scarcely on FID.
In most cases, the utilisation of the frequency loss has a positive impact on reconstruction quality \textcolor{black}{metrics} -- both SSIM/PSNR and FID.

For SwinMR (PI), the utilisation of perceptual loss tends to slightly decrease SSIM and PSNR, but substantially decreases the FID (PFP vs PF; PP vs P).
For SwinMR (nPI), the utilisation of perceptual loss tends to achieve a better FID but scarcely change SSIM and PSNR (PFP vs PF; PP vs P).
In most cases, the utilisation of the perceptual loss has a positive impact on FID, but a negative impact on SSIM/PSNR when using multi-channel data.

\subsection{Downstream Task Experiments: Brain Segmentation Experiments on BraTS17 \textcolor{black}{Dataset}}

In this experiment, we performed a downstream task using a reconstructed MR image, in order to measure the reconstruction quality.
\textcolor{black}{Specifically, we chose an open-access multi-modalities brain tumour segmentation network\footnote{https://github.com/Mehrdad-Noori/Brain-Tumor-Segmentation}~\citep{Noori2019} for the downstream task experiments. 
This segmentation network adopted a U-Net~\citep{Ronneberger2015} based architecture with the utilisation of residual blocks and strided convolution downsampling compared to the vanilla U-Net. In addition, this segmentation network also employed the Squeeze-and-Excitation Block~\citep{Hu2017} on concatenated multi-level features for channel attention mechanism.}

\textcolor{black}{The segmentation network was trained on the BraTS17 dataset (four modalities are required including FLAIR, T1, T1CE and T2).}
Then, we trained four SwinMR weights using BraTS17 FLAIR, T1, T1CE and T2 data, respectively. After that, segmentation tasks were conducted on GT MR images, SwinMR reconstructed MR images and ZF MR images directly using the pre-trained segmentation network. Ideally, the segmentation score of reconstructed images and GT images should be as closer as possible.

\begin{figure}[H]
    \centering
    \includegraphics[width=5in]{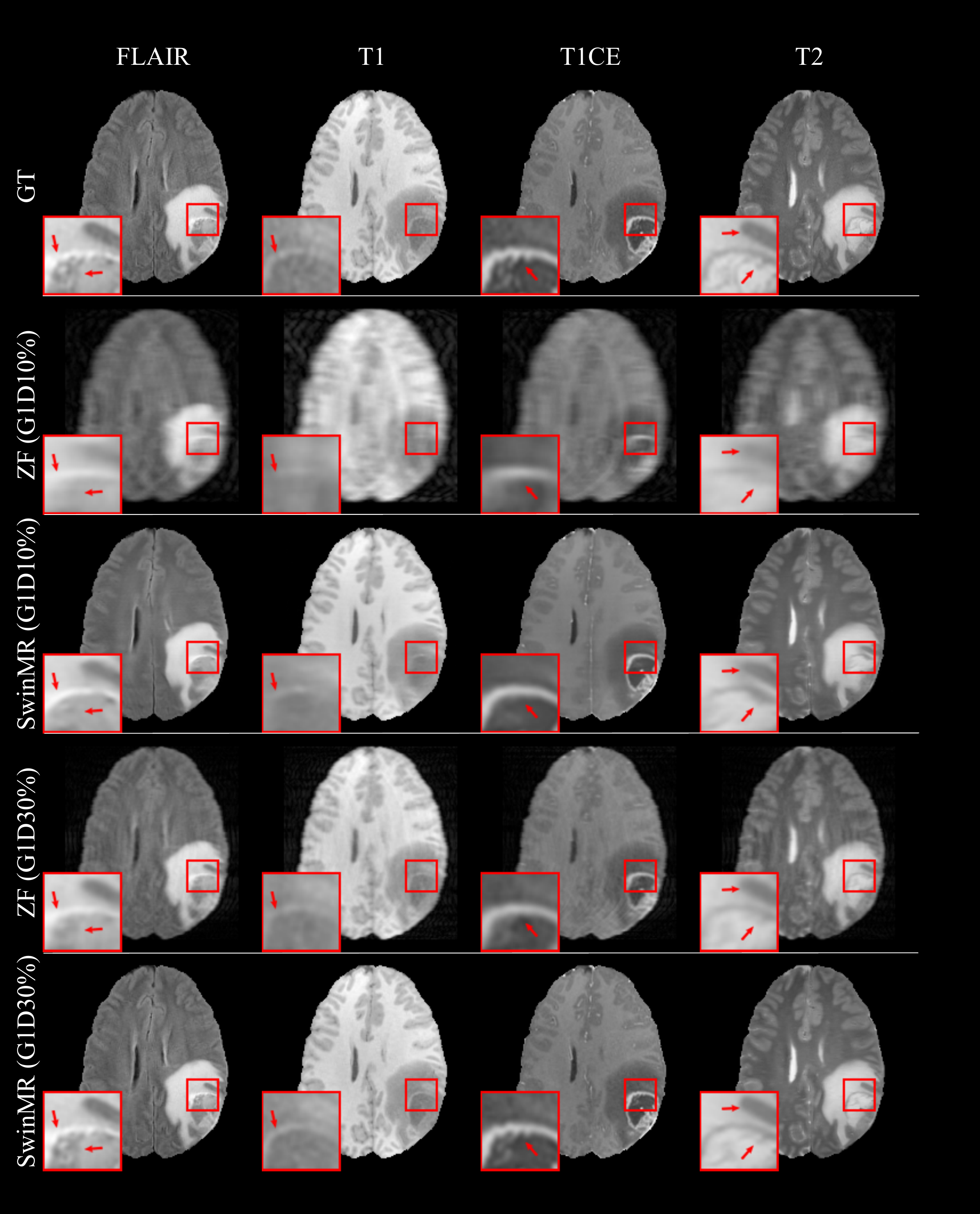}
    \caption{
    Samples of reconstruction results for SwinMR on BraTS17 dataset including FLAIR, T1, T1CE and T2 MR images.
    Row 1: Ground truth MR images (GT);
    Row 2: Zero-filled MR images (ZF) undersampled by Gaussian 1D 10\% mask (G1D10\%);
    Row 3: Reconstructed MR images undersampled by G1D10\%;
    Row 4: ZF undersampled by Gaussian 1D 30\% mask (G1D30\%);
    Row 5: Reconstructed MR images undersampled by G1D30\%.
    }
    \label{fig:FIG_EXP_IMAGE_BraTS17_Reconstruction}
\end{figure}

\begin{figure}[H]
    \centering
    \includegraphics[width=5in]{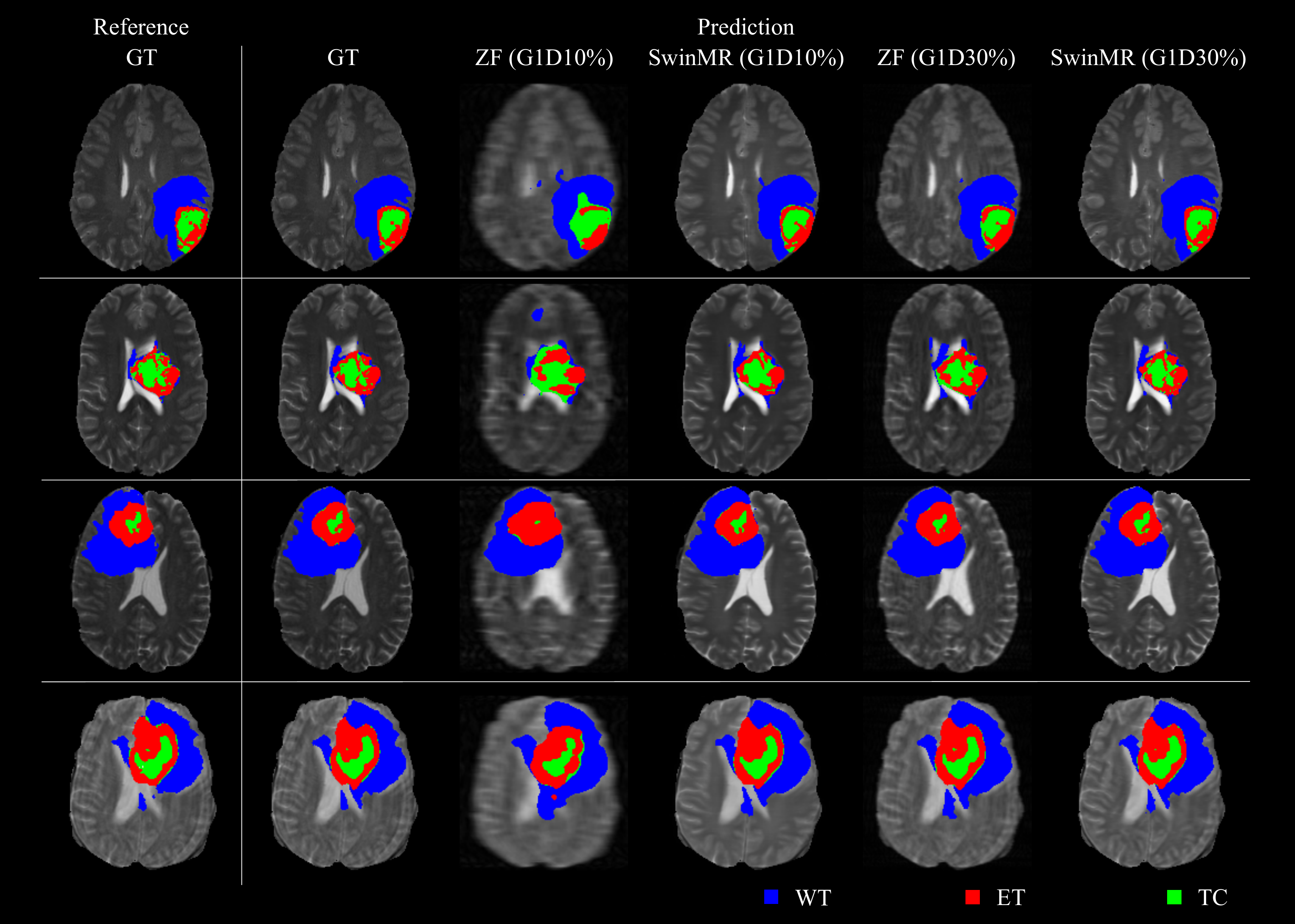}
    \caption{
    Samples of segmentation results for SwinMR on the BraTS17 dataset.
    Col 1: Segmentation reference;
    Col 2: Segmentation prediction using GT images;
    Col 3: Segmentation prediction using zero-filled MR images (ZF) undersampled by Gaussian 1D 10\% mask (G1D10\%);
    Col 4: Segmentation prediction using reconstructed MR images undersampled by G1D10\%;
    Col 5: Segmentation prediction using ZF undersampled by Gaussian 1D 30\% mask (G1D30\%);
    Col 6: Segmentation prediction using reconstructed MR images undersampled by G1D30\%.
    Blue area: Whole tumour (WT);
    Red area: Enhancing tumour (ET);
    Green area: Tumour core (TC).
    }
    \label{fig:FIG_EXP_IMAGE_BraTS17_Segmentation}
\end{figure}

Table~\ref{tab:BraTS17_Recon} shows the result of SwinMR trained with BraTS17 FLAIR, T1, T1CE and T2 respectively.
Figure~\ref{fig:FIG_EXP_IMAGE_BraTS17_Reconstruction} displays the samples of the reconstruction of different modalities.
Table~\ref{tab:BraTS17_IoU} and Table~\ref{tab:BraTS17_Dice} show the IoU and Dice score of the segmentation task.
Figure~\ref{fig:FIG_EXP_IMAGE_BraTS17_Segmentation} displays the sample of the segmentation task.

According to Table~\ref{tab:BraTS17_IoU} and Table~\ref{tab:BraTS17_Dice}, the IoU and Dice score of reconstructed MR images are improved compared with ZF MR images and much closer to the score of GT MR images. 
According to the Mann-Whitney Test, the IoU and Dice score distributions of the reconstructed MR images using the Gaussian 1D 30\% mask are not significantly different from the distributions of the GT MR images ($p > 0.05$).

\begin{table}[H]
    \centering
    \caption{
    Quantitative results of reconstructed images by SwinMR (Recon) and zero-filled images (ZF) on BraTS17 dataset (mean (std)).
    PSNR: Peak signal-to-noise ratio;
    SSIM: Structural similarity index;
    FID: Fr\'echet inception distance.
    G1D10\%: Gaussian 1D 10\% mask;
    G1D30\%: Gaussian 1D 30\% mask.\\
    }
    \scalebox{0.75}{
    \begin{tabular}{cccccc}
        \toprule
        \multirow{2}[4]{*}{Mask} & \multirow{2}[4]{*}{Metrics} & \multicolumn{4}{c}{Recon} \\
    \cmidrule{3-6}          &       & FLAIR & T1    & T1CE  & T2 \\
        \midrule
        \multirow{3}[2]{*}{G1D10\%} & PSNR  & 30.07 (1.99) & 33.80 (2.30) & 33.80 (1.84) & 32.20 (1.81) \\
              & SSIM  & 0.751 (0.043) & 0.760 (0.046) & 0.797 (0.049) & 0.745 (0.039) \\
              & FID   & 38.02 & 32.97 & 31.46 & 21.84 \\
        \midrule
        \multirow{3}[2]{*}{G1D30\%} & PSNR  & 37.97 (2.42) & 41.08 (3.36) & 42.29 (2.12) & 38.37 (2.02) \\
              & SSIM  & 0.942 (0.013) & 0.953 (0.012) & 0.953 (0.015) & 0.937 (0.016) \\
              & FID   & 5.94  & 4.80  & 4.39  & 8.95 \\
        \midrule
        \multirow{2}[4]{*}{Mask} & \multirow{2}[4]{*}{Metrics} & \multicolumn{4}{c}{ZF} \\
    \cmidrule{3-6}          &       & FLAIR & T1    & T1CE  & T2 \\
        \midrule
        \multirow{3}[2]{*}{G1D10\%} & PSNR  & 23.87 (1.64) & 25.92 (1.48) & 25.92 (1.70) & 23.92 (1.79) \\
              & SSIM  & 0.388 (0.070) & 0.414 (0.061) & 0.414 (0.068) & 0.431 (0.057) \\
              & FID   & 225.70 & 234.52 & 227.51 & 219.09 \\
        \midrule
        \multirow{3}[2]{*}{G1D30\%} & PSNR  & 28.74 (1.78) & 28.82 (1.60) & 30.60 (1.82) & 29.46 (2.01) \\
              & SSIM  & 0.597 (0.046) & 0.602 (0.051) & 0.602 (0.051) & 0.632 (0.038) \\
              & FID   & 91.18 & 100.98 & 106.28 & 85.49 \\
        \bottomrule
    \end{tabular}}%
    \label{tab:BraTS17_Recon}%
\end{table}%

\begin{table}[H]
    \centering
    \caption{
    Intersection over union (IoU) of the segmentation experiment (median/mean [Q1,Q3]). 
    $^{\star}$: $p < 0.05$;
    $^{\star\star}$: $p < 0.01$
    (compared with GT by Mann-Whitney Test).
    GT: ground truth MR images; Recon: reconstructed MR images by SwinMR; ZF: undersampled zero-filled MR images.
    G1D10\%: Gaussian 1D 10\% mask; G1D30\%: Gaussian 1D 30\% mask.
    WT: Whole tumour; TC: Enhancing tumour ; ET: Tumour core.\\
    }
    \scalebox{0.75}{
    \begin{tabular}{ccccc}
    \toprule
    \multicolumn{2}{c}{IoU} & GT    & Recon & ZF \\
    \midrule
    \multirow{3}[2]{*}{G1D10\%} & WT    & 0.930/0.924 [0.900,0.954] & 0.898/0.899 [0.868,0.940]$^{\star\star}$ & 0.838/0.836 [0.795,0.881]$^{\star\star}$ \\
          & TC    & 0.821/0.771 [0.726,0.903] & 0.758/0.722 [0.661,0.890]$^{\star\star}$ & 0.617/0.539 [0.393,0.733]$^{\star\star}$ \\
          & ET    & 0.772/0.735 [0.625,0.889] & 0.740/0.652 [0.471,0.846]$^{\star\star}$ & 0.570/0.527 [0.336,0.694]$^{\star\star}$ \\
    \midrule
    \multirow{3}[2]{*}{G1D30\%} & WT    & 0.930/0.924 [0.900,0.954] & 0.924/0.921 [0.895,0.953] & 0.897/0.897 [0.862,0.945]$^{\star\star}$ \\
          & TC    & 0.821/0.771 [0.726,0.903] & 0.811/0.766 [0.719,0.904] & 0.763/0.728 [0.669,0.895]$^{\star\star}$ \\
          & ET    & 0.772/0.735 [0.625,0.889] & 0.770/0.725 [0.616,0.883] & 0.748/0.697 [0.573,0.859]$^{\star\star}$ \\
    \bottomrule
    \end{tabular}}%
    \label{tab:BraTS17_IoU}%
\end{table}%

\begin{table}[H]
    \centering
    \caption{
    Dice score of the segmentation experiment (median/mean [Q1,Q3]). 
    $^{\star}$: $p < 0.05$;
    $^{\star\star}$: $p < 0.01$
    (compared with GT by Mann-Whitney Test).
    GT: ground truth MR images; Recon: reconstructed MR images by SwinMR; ZF: undersampled zero-filled MR images.
    G1D10\%: Gaussian 1D 10\% mask; G1D30\%: Gaussian 1D 30\% mask.
    WT: Whole tumour; TC: Enhancing tumour; ET: Tumour core.\\
    }
    \scalebox{0.75}{
    \begin{tabular}{ccccc}
    \toprule
    \multicolumn{2}{c}{Dice} & GT    & Recon & ZF \\
    \midrule
    \multirow{3}[2]{*}{G1D10\%} & WT    & 0.968/0.965 [0.952,0.981] & 0.950/0.950 [0.933,0.974]$^{\star\star}$ & 0.916/0.914 [0.892,0.940]$^{\star\star}$ \\
          & TC    & 0.904/0.857 [0.845,0.951] & 0.863/0.819 [0.800,0.944]$^{\star\star}$ & 0.767/0.653 [0.566,0.847]$^{\star\star}$ \\
          & ET    & 0.874/0.835 [0.777,0.941] & 0.852/0.766 [0.640,0.917]$^{\star\star}$ & 0.725/0.665 [0.503,0.820]$^{\star\star}$ \\
    \midrule
    \multirow{3}[2]{*}{G1D30\%} & WT    & 0.968/0.965 [0.952,0.981] & 0.964/0.963 [0.948,0.980] & 0.949/0.949 [0.930,0.975]$^{\star\star}$ \\
          & TC    & 0.904/0.857 [0.845,0.951] & 0.897/0.854 [0.838,0.951] & 0.868/0.826 [0.803,0.947]$^{\star\star}$ \\
          & ET    & 0.874/0.835 [0.777,0.941] & 0.871/0.827 [0.765,0.939] & 0.857/0.808 [0.729,0.925]$^{\star\star}$ \\
    \bottomrule
    \end{tabular}}%
    \label{tab:BraTS17_Dice}%
\end{table}%

\section{Discussion}
% overall
In this work, a novel Swin transformer based model, i.e., SwinMR, for fast MRI reconstruction \textcolor{black}{has} been proposed. 
Most existing deep learning based image restoration methods, including MRI reconstruction approaches, are based on CNNs. The convolution is a very effective feature extractor but lacks long-range dependency. The receptive field of CNNs is limited by the size of the kernel and the depth of the network. 
To tackle this problem, researchers have developed transformers based image restoration methods that have been originally used for solving NLP tasks. The core of the transformer is MSA, which has global sensitivity. In MSA operation, each patch can link with any other patches in the whole image space but also aggravates the computational burden.

% Why Swin Transformer?
% Here comes a question: do MRI images need the feature extractor with global sensitivity? 
However, we have believed that in MRI reconstruction, the MSA, which is operated in the whole image space, is redundant and not necessary.
It is not difficult to understand that in NLP tasks the first and the last words may have a strong connection in a sentence. However, this may not be applicable in CV tasks. 
Visual elements (e.g., pixels) in CV tasks can vary substantially in scale unlike language elements (e.g., word tokens) in NLP tasks~\citep{Liu2021}. 
Since in most cases, for example, the top-left corner patch has no relationship with the bottom-right corner patch within an image. Moreover, for MRI reconstruction, the biggest difficulty is the recovery of detailed information and texture information. Focusing too much on global information and ignoring the detailed (local) information may make the image smoother and lose more details.
The utilisation of a Swin transformer can achieve a trade-off for CV tasks. In Swin transformer, operations are conducted in shifted windows instead of the whole images. It has a larger receptive field compared to CNNs but is not overly concerned with global information. This is the reason why we have developed a Swin transformer for MRI reconstruction.

% Analyse of Comparison
To evaluate our proposed methods, several comparison experiments and ablation studies have been conducted. In this study, we have compared our proposed SwinMR with benchmark MRI reconstruction methods. The results in Table~\ref{tab:comparison} have demonstrated that our SwinMR has achieved the highest SSIM/PSNR and lowest FID compared to CNN-based and GAN-based models. From Figure~\ref{fig:FIG_EXP_IMAGE_Comparison}, we have shown clearly that our SwinMR has obtained better reconstruction quality, especially in the zoom-in area, where the details of the cerebellum have been well-preserved.

% PI vs nPI
In this study, we have also compared SwinMR (PI) that has been trained with multi-channel brain data with SwinMR (nPI) that has been trained with single-channel brain data. The results have led to a similar conclusion in our previous study~\citep{Huang2021}, where FID of the model trained with multi-channel data has been better compared to the model trained with single-channel data, and the SSIM/PSNR has shown the opposite (i.e., SSIM/PSNR: nPI \textgreater PI; FID: PI \textless nPI). This phenomenon can also be observed in the subsequent 
ablation experiments.
From Figure~\ref{fig:FIG_EXP_IMAGE_Comparison}, we can find that the reconstructed images of SwinMR (PI) have shown more details and texture information, but the reconstructed images of SwinMR (nPI) have shown smoother.

%So why do these three metrics that compared PI and nPI give a different answer? 
The experimental results have demonstrated that the three metrics that compared PI and nPI gave different answers. We have speculated that this might be due to the different principles of these metrics.
PSNR is a classic metric based on per-pixel comparisons, which are not able to reflect the structure information for images. SSIM is a perceptual metric that measures structure similarity. However, both of them are based on simple and shallow functions and direct comparisons between images, which is insufficient to account for many nuances of human perception~\citep{Zhang2018}.
For FID, the comparison is based on perception and performed on two sets of images. Images are mapped to high-dimension representations by a pre-trained InceptionV3 network, which is well-related to human visual perception.
The SwinMR (PI) reconstructed images have demonstrated more details and texture information. Even though these details and texture information may not be so \emph{accurate}, they make the reconstructed images more \emph{visually similar} with the ground truth images.
However, the SwinMR (nPI) reconstructed images have shown smoother in pixel-wise scale, at the cost of less detail and texture information. 
Therefore, SwinMR (PI) have tended to have better FID and worse SSIM/PSNR compared to SwinMR (nPI), due to the principle differences of the evaluation methods.

% Time cost
From Table~\ref{tab:comparison}, we can find a common problem of transformer-based methods, which is the higher computational cost compared to other CNN-based and GAN-based methods. Equation (\ref{formula:11}) have shown that the computational complexity is proportional to the $HW$ of the input of (S)W-MSA. The time shown in Table~\ref{tab:comparison} has been the inference time, where the original height and weight have been treated as $H$ and $W$ ($256 \times 256$ here). For training, randomly cropping have been applied to ease the long processing time.

% Mask and Noise
Experiments using different undersampling masks with various noise levels have demonstrated that our proposed method SwinMR have shown superiority to DAGAN in all the tests. The evaluation \textcolor{black}{metrics} change as expected when the condition changes (different masks and noise levels). 

% Patch and Channel
Ablation studies on the patch number and the channel number have demonstrated that reconstruction quality has been improved as the patch number has been increased and has gradually been saturated, according to Figures~\ref{fig:FIG_AEXP_GRAPH_ChannelPatch}(A) and (C).
However, according to Equation (\ref{formula:11}), the computational complexity also has been increased as the patch number has been increased. As a trade-off, we have set the patch number to 96.
Beyond our expectations, the changing of channel number has not been positively correlated with the evaluation \textcolor{black}{metrics} in this experiment, according to Figures~\ref{fig:FIG_AEXP_GRAPH_ChannelPatch}(B) and (D). We have assumed that the evaluation \textcolor{black}{metrics} have saturated in the range of channel number in this experiment. Empirically, we have set the channel number to 180 according to the default setting of SwinIR. 

Ablation studies on different loss functions have been conducted. As expected, the utilisation of the pixel-wise loss and the frequency loss has mainly constrained the fidelity of reconstruction, and the utilisation of perceptual VGG loss has focused on perception, which has been well-related to the human visual system. Therefore, the utilisation of frequency loss has had a positive impact on SSIM and PSNR, which has been more sensitive to the fidelity of reconstruction. The utilisation of perceptual loss has had a positive impact on FID, which has been based on perception.

There are still some limitations of our work. 
First, in the (S)W-MSA operation, the size of windows is fixed. Inspired by Google-Net, multi-scale windows could be incorporated and results from different scales could be merged in the (S)W-MSA.
Second, the heavy computational cost is still an obstacle to the development of transformers. The improvement that transformers bring is at the sacrifice of increased computational cost. A lightweight transformer model could be a potential future research direction.

\section{Conclusion}

In this work, we have developed the SwinMR, a novel parallel imaging coupled Swin transformer-based model for fast multi-channel MRI reconstruction. The proposed method has outperformed other benchmark CNN-based and GAN-based MRI reconstruction methods. It has also shown excellent robustness using different undersampling trajectories with various noises.

\section*{Acknowledgement}

This work was supported in part by the UK Research and Innovation Future Leaders Fellowship [MR/V023799/1], in part by the Medical Research Council [MC/PC/21013], in part by the European Research Council Innovative Medicines Initiative [DRAGON, H2020-JTI-IMI2 101005122], in part by the AI for Health Imaging Award [CHAIMELEON, H2020-SC1-FA-DTS-2019-1 952172], in part by the British Heart Foundation [Project Number: TG/18/5/34111, PG/16/78/32402], in part by the Project of Shenzhen International Cooperation Foundation [GJHZ20180926165402083], in part by the Basque Government through the ELKARTEK funding program [KK-2020/00049], in part by the consolidated research group MATHMODE [IT1294-19], and in part by the NVIDIA Academic Hardware Grants.

\newpage


\bibliographystyle{elsarticle-num} 
\bibliography{references.bib}

\begin{thebibliography}{10}
\expandafter\ifx\csname url\endcsname\relax
  \def\url#1{\texttt{#1}}\fi
\expandafter\ifx\csname urlprefix\endcsname\relax\def\urlprefix{URL }\fi
\expandafter\ifx\csname href\endcsname\relax
  \def\href#1#2{#2} \def\path#1{#1}\fi

\bibitem{Zbontar2018}
J.~{Zbontar}, F.~{Knoll}, A.~{Sriram}, T.~{Murrell}, Z.~{Huang}, M.~J.
  {Muckley}, A.~{Defazio}, R.~{Stern}, P.~{Johnson}, M.~{Bruno}, M.~{Parente},
  K.~J. {Geras}, J.~{Katsnelson}, H.~{Chandarana}, Z.~{Zhang}, M.~{Drozdzal},
  A.~{Romero}, M.~{Rabbat}, P.~{Vincent}, N.~{Yakubova}, J.~{Pinkerton},
  D.~{Wang}, E.~{Owens}, C.~L. {Zitnick}, M.~P. {Recht}, D.~K. {Sodickson},
  Y.~W. {Lui}, {FastMRI}: An open dataset and benchmarks for accelerated {MRI},
  arXiv e-prints (2018) arXiv:1811.08839.

\bibitem{Stehling1991}
M.~K. Stehling, R.~Turner, P.~Mansfield, {Echo-planar} imaging: Magnetic
  resonance imaging in a fraction of a second, Science 254~(5028) (1991)
  43--50.

\bibitem{Hennig1986}
J.~Hennig, A.~Nauerth, H.~Friedburg, {RARE} imaging: A fast imaging method for
  clinical {MR}, Magnetic Resonance in Medicine 3~(6) (1986) 823--833.
\newblock \href {https://doi.org/https://doi.org/10.1002/mrm.1910030602}
  {\path{doi:https://doi.org/10.1002/mrm.1910030602}}.

\bibitem{Blaimer2004}
M.~Blaimer, F.~Breuer, M.~Mueller, R.~M. Heidemann, M.~A. Griswold, P.~M.
  Jakob, {SMASH}, {SENSE}, {PILS}, {GRAPPA}: How to choose the optimal method,
  Topics in Magnetic Resonance Imaging 15~(4) (2004) 223--236.

\bibitem{Sodickson1997}
D.~K. Sodickson, W.~J. Manning, Simultaneous acquisition of spatial harmonics
  ({SMASH}): Fast imaging with radiofrequency coil arrays, Magnetic Resonance
  in Medicine 38~(4) (1997) 591--603.

\bibitem{Pruessmann1999}
K.~P. Pruessmann, M.~Weiger, M.~B. Scheidegger, P.~Boesiger, {SENSE}:
  Sensitivity encoding for fast {MRI}, Magnetic Resonance in Medicine: An
  Official Journal of the International Society for Magnetic Resonance in
  Medicine 42~(5) (1999) 952--962.

\bibitem{Griswold2002}
M.~A. Griswold, P.~M. Jakob, R.~M. Heidemann, M.~Nittka, V.~Jellus, J.~Wang,
  B.~Kiefer, A.~Haase, Generalized autocalibrating partially parallel
  acquisitions ({GRAPPA}), Magnetic Resonance in Medicine 47 (2002) 1202--1210.
\newblock \href {https://doi.org/10.1002/mrm.10171}
  {\path{doi:10.1002/mrm.10171}}.

\bibitem{Donoho2006}
D.~Donoho, Compressed sensing, IEEE Transactions on Information Theory 52
  (2006) 1289--1306.
\newblock \href {https://doi.org/10.1109/TIT.2006.871582}
  {\path{doi:10.1109/TIT.2006.871582}}.

\bibitem{Block2007}
K.~T. Block, M.~Uecker, J.~Frahm, Undersampled radial {MRI} with multiple
  coils. {Iterative} image reconstruction using a total variation constraint,
  Magnetic Resonance in Medicine: An Official Journal of the International
  Society for Magnetic Resonance in Medicine 57~(6) (2007) 1086--1098.

\bibitem{Beladgham2008}
M.~Beladgham, I.~B. Hacene, A.~Taleb-Ahmed, M.~Kh{\'e}lif, {MRI} images
  compression using curvelets transforms, in: AIP Conference Proceedings, Vol.
  1019, American Institute of Physics, 2008, pp. 249--253.

\bibitem{Zhu2013}
Z.~Zhu, K.~Wahid, P.~Babyn, R.~Yang, Compressed sensing-based {MRI}
  reconstruction using complex double-density dual-tree {DWT}, Journal of
  Biomedical Imaging 2013 (jan 2013).
\newblock \href {https://doi.org/10.1155/2013/907501}
  {\path{doi:10.1155/2013/907501}}.

\bibitem{Ravishankar2011}
S.~Ravishankar, Y.~Bresler, {MR} image reconstruction from highly undersampled
  k-space data by dictionary learning, IEEE Transactions on Medical Imaging
  30~(5) (2011) 1028--1041.
\newblock \href {https://doi.org/10.1109/TMI.2010.2090538}
  {\path{doi:10.1109/TMI.2010.2090538}}.

\bibitem{Zeng2019}
N.~Zeng, Z.~Wang, H.~Zhang, K.-E. Kim, Y.~Li, X.~Liu, An improved particle
  filter with a novel hybrid proposal distribution for quantitative analysis of
  gold immunochromatographic strips, IEEE Transactions on Nanotechnology 18
  (2019) 819--829.
\newblock \href {https://doi.org/10.1109/TNANO.2019.2932271}
  {\path{doi:10.1109/TNANO.2019.2932271}}.

\bibitem{Zeng2021}
N.~Zeng, H.~Li, Y.~Peng, A new deep belief network-based multi-task learning
  for diagnosis of {Alzheimer's} disease, Neural Computing and Applications
  (2021).

\bibitem{Wu2022}
P.~Wu, H.~Li, N.~Zeng, F.~Li, {FMD-Yolo}: An efficient face mask detection
  method for {COVID-19} prevention and control in public, Image and Vision
  Computing 117 (2022) 104341.

\bibitem{Chen2022}
Y.~Chen, C.-B. Sch{\"o}nlieb, P.~Li{\`o}, T.~Leiner, P.~L. Dragotti, G.~Wang,
  D.~Rueckert, D.~Firmin, G.~Yang, {AI}-based reconstruction for fast {MRI}-a
  systematic review and meta-analysis, Proceedings of the IEEE 110~(2) (2022)
  224--245.

\bibitem{Bakator2018}
M.~Bakator, D.~Radosav, Deep learning and medical diagnosis: A review of
  literature, Multimodal Technologies and Interaction 2~(3) (2018).

\bibitem{Girshick2014}
R.~Girshick, J.~Donahue, T.~Darrell, J.~Malik, Rich feature hierarchies for
  accurate object detection and semantic segmentation, in: Proceedings of the
  IEEE Conference on Computer Vision and Pattern Recognition (CVPR), 2014.

\bibitem{Szegedy2015}
C.~Szegedy, W.~Liu, Y.~Jia, P.~Sermanet, S.~Reed, D.~Anguelov, D.~Erhan,
  V.~Vanhoucke, A.~Rabinovich, Going deeper with convolutions, in: Proceedings
  of the IEEE Conference on Computer Vision and Pattern Recognition (CVPR),
  2015.

\bibitem{Long2015}
J.~Long, E.~Shelhamer, T.~Darrell, Fully convolutional networks for semantic
  segmentation, in: Proceedings of the IEEE Conference on Computer Vision and
  Pattern Recognition (CVPR), 2015.

\bibitem{Dong2016}
C.~Dong, C.~C. Loy, K.~He, X.~Tang, Image super-resolution using deep
  convolutional networks, IEEE Transactions on Pattern Analysis and Machine
  Intelligence 38 (2016) 295--307.
\newblock \href {https://doi.org/10.1109/TPAMI.2015.2439281}
  {\path{doi:10.1109/TPAMI.2015.2439281}}.

\bibitem{Wang2016}
S.~Wang, Z.~Su, L.~Ying, X.~Peng, S.~Zhu, F.~Liang, D.~Feng, D.~Liang,
  Accelerating magnetic resonance imaging via deep learning, in: 2016 IEEE 13th
  International Symposium on Biomedical Imaging (ISBI), 2016, pp. 514--517.
\newblock \href {https://doi.org/10.1109/ISBI.2016.7493320}
  {\path{doi:10.1109/ISBI.2016.7493320}}.

\bibitem{Yang2016}
Y.~Yang, J.~Sun, H.~Li, Z.~Xu, Deep {ADMM-Net} for compressive sensing {MRI},
  in: Advances in Neural Information Processing Systems, Vol.~29, Curran
  Associates, Inc., 2016.

\bibitem{Schlemper2018}
J.~Schlemper, J.~Caballero, J.~V. Hajnal, A.~N. Price, D.~Rueckert, A deep
  cascade of convolutional neural networks for dynamic {MR} image
  reconstruction, IEEE Transactions on Medical Imaging 37 (2018) 491--503.
\newblock \href {https://doi.org/10.1109/TMI.2017.2760978}
  {\path{doi:10.1109/TMI.2017.2760978}}.

\bibitem{Zhu2018}
B.~Zhu, J.~Z. Liu, S.~F. Cauley, B.~R. Rosen, M.~S. Rosen, Image reconstruction
  by domain-transform manifold learning, Nature 555 (2018) 487--492.
\newblock \href {https://doi.org/10.1038/nature25988}
  {\path{doi:10.1038/nature25988}}.

\bibitem{Sutskever2014}
I.~Sutskever, O.~Vinyals, Q.~V. Le, Sequence to sequence learning with neural
  networks, in: Advances in Neural Information Processing Systems, Vol.~27,
  Curran Associates, Inc., 2014.

\bibitem{Vaswani2017}
A.~Vaswani, N.~Shazeer, N.~Parmar, J.~Uszkoreit, L.~Jones, A.~N. Gomez,
  {\L}.~Kaiser, I.~Polosukhin, Attention is all you need, in: Advances in
  neural information processing systems, 2017, pp. 5998--6008.

\bibitem{Parikh2016}
A.~P. {Parikh}, O.~{T{\"a}ckstr{\"o}m}, D.~{Das}, J.~{Uszkoreit}, {A
  Decomposable Attention Model for Natural Language Inference}, arXiv e-prints
  (2016) arXiv:1606.01933.

\bibitem{Cheng2016}
J.~{Cheng}, L.~{Dong}, M.~{Lapata}, {Long Short-Term Memory-Networks for
  Machine Reading}, arXiv e-prints (2016) arXiv:1601.06733.

\bibitem{Matsoukas2021}
C.~{Matsoukas}, J.~{Fredin Haslum}, M.~{S{\"o}derberg}, K.~{Smith}, Is it time
  to replace {CNNs} with transformers for medical images?, arXiv e-prints
  (2021) arXiv:2108.09038.

\bibitem{Parmar2018}
N.~Parmar, A.~Vaswani, J.~Uszkoreit, L.~Kaiser, N.~Shazeer, A.~Ku, D.~Tran,
  Image transformer, in: Proceedings of the 35th International Conference on
  Machine Learning, Vol.~80 of Proceedings of Machine Learning Research, PMLR,
  2018, pp. 4055--4064.

\bibitem{Salimans2017}
T.~{Salimans}, A.~{Karpathy}, X.~{Chen}, D.~P. {Kingma}, {PixelCNN++}:
  Improving the {PixelCNN} with discretized logistic mixture likelihood and
  other modifications, arXiv e-prints (2017) arXiv:1701.05517.

\bibitem{Qiu2020}
X.~Qiu, T.~Sun, Y.~Xu, Y.~Shao, N.~Dai, X.~Huang, Pre-trained models for
  natural language processing: A survey, Science China Technological Sciences
  (2020) 1--26.

\bibitem{Carion2020}
N.~Carion, F.~Massa, G.~Synnaeve, N.~Usunier, A.~Kirillov, S.~Zagoruyko,
  End-to-end object detection with transformers, in: Computer Vision -- ECCV
  2020, Springer International Publishing, Cham, 2020, pp. 213--229.

\bibitem{Dosovitskiy2020}
A.~{Dosovitskiy}, L.~{Beyer}, A.~{Kolesnikov}, D.~{Weissenborn}, X.~{Zhai},
  T.~{Unterthiner}, M.~{Dehghani}, M.~{Minderer}, G.~{Heigold}, S.~{Gelly},
  J.~{Uszkoreit}, N.~{Houlsby}, An image is worth {16x16} words: Transformers
  for image recognition at scale, arXiv e-prints (2020) arXiv:2010.11929.

\bibitem{Liu2021}
Z.~{Liu}, Y.~{Lin}, Y.~{Cao}, H.~{Hu}, Y.~{Wei}, Z.~{Zhang}, S.~{Lin},
  B.~{Guo}, {Swin Transformer: Hierarchical Vision Transformer using Shifted
  Windows}, arXiv e-prints (2021) arXiv:2103.14030.

\bibitem{Yang2018}
G.~Yang, S.~Yu, H.~Dong, G.~Slabaugh, P.~L. Dragotti, X.~Ye, F.~Liu,
  S.~Arridge, J.~Keegan, Y.~Guo, D.~Firmin, {DAGAN}: Deep de-aliasing
  generative adversarial networks for fast compressed sensing {MRI}
  reconstruction, IEEE Transactions on Medical Imaging 37 (2018) 1310--1321.
\newblock \href {https://doi.org/10.1109/TMI.2017.2785879}
  {\path{doi:10.1109/TMI.2017.2785879}}.

\bibitem{Shin2020}
H.-C. {Shin}, A.~{Ihsani}, S.~{Mandava}, S.~{Turuvekere Sreenivas},
  C.~{Forster}, J.~{Cha}, A.~{Disease Neuroimaging Initiative}, {GANBERT}:
  Generative adversarial networks with bidirectional encoder representations
  from transformers for {MRI} to {PET} synthesis, arXiv e-prints (2020)
  arXiv:2008.04393.

\bibitem{Zhang2021}
X.~{Zhang}, X.~{He}, J.~{Guo}, N.~{Ettehadi}, N.~{Aw}, D.~{Semanek},
  J.~{Posner}, A.~{Laine}, Y.~{Wang}, {PTNet}: A high-resolution infant {MRI}
  synthesizer based on transformer, arXiv e-prints (2021) arXiv:2105.13993.

\bibitem{Dalmaz2021}
O.~{Dalmaz}, M.~{Yurt}, T.~{{\c{C}}ukur}, {ResViT}: Residual vision
  transformers for multi-modal medical image synthesis, arXiv e-prints (2021)
  arXiv:2106.16031.

\bibitem{Korkmaz2021_1}
Y.~Korkmaz, M.~Yurt, S.~U.~H. Dar, T.~Cukur, Deep {MRI} reconstruction with
  generative vision transformers, in: Machine Learning for Medical Image
  Reconstruction, Springer International Publishing, Cham, 2021, pp. 54--64.

\bibitem{Korkmaz2021_2}
Y.~Korkmaz, S.~U. Dar, M.~Yurt, M.~Özbey, T.~Çukur, Unsupervised {MRI}
  reconstruction via zero-shot learned adversarial transformers, IEEE
  Transactions on Medical Imaging (2022) 1--1\href
  {https://doi.org/10.1109/TMI.2022.3147426}
  {\path{doi:10.1109/TMI.2022.3147426}}.

\bibitem{Feng2021_1}
C.-M. Feng, Y.~Yan, H.~Fu, L.~Chen, Y.~Xu, Task transformer network for joint
  {MRI} reconstruction and super-resolution, in: Medical Image Computing and
  Computer Assisted Intervention -- MICCAI 2021, Springer International
  Publishing, Cham, 2021, pp. 307--317.

\bibitem{Feng2021_2}
C.-M. {Feng}, Y.~{Yan}, G.~{Chen}, H.~{Fu}, Y.~{Xu}, L.~{Shao}, Accelerated
  multi-modal {MR} imaging with transformers, arXiv e-prints (2021)
  arXiv:2106.14248.

\bibitem{Liang2021}
J.~Liang, J.~Cao, G.~Sun, K.~Zhang, L.~Van~Gool, R.~Timofte, {SwinIR}: Image
  restoration using swin transformer, in: Proceedings of the IEEE/CVF
  International Conference on Computer Vision (ICCV) Workshops, 2021, pp.
  1833--1844.

\bibitem{qu2012undersampled}
X.~Qu, D.~Guo, B.~Ning, Y.~Hou, Y.~Lin, S.~Cai, Z.~Chen, Undersampled {MRI}
  reconstruction with patch-based directional wavelets, Magnetic resonance
  imaging 30~(7) (2012) 964--977.

\bibitem{wu2017solving}
T.~Wu, D.~Z. Wang, Z.~Jin, J.~Zhang, Solving constrained {TV2L1-L2} {MRI}
  signal reconstruction via an efficient alternating direction method of
  multipliers, Numerical Mathematics: Theory, Methods and Applications 10~(4)
  (2017) 895--912.

\bibitem{ravishankar2010mr}
S.~Ravishankar, Y.~Bresler, {MR} image reconstruction from highly undersampled
  k-space data by dictionary learning, IEEE transactions on medical imaging
  30~(5) (2010) 1028--1041.

\bibitem{lustig2007sparse}
M.~Lustig, D.~Donoho, J.~M. Pauly, Sparse {MRI}: The application of compressed
  sensing for rapid {MR} imaging, Magnetic Resonance in Medicine: An Official
  Journal of the International Society for Magnetic Resonance in Medicine
  58~(6) (2007) 1182--1195.

\bibitem{yang2010fast}
J.~Yang, Y.~Zhang, W.~Yin, A fast alternating direction method for {TVL1-L2}
  signal reconstruction from partial fourier data, IEEE Journal of Selected
  Topics in Signal Processing 4~(2) (2010) 288--297.

\bibitem{wang2014compressed}
L.~Wang, K.~Lu, P.~Liu, Compressed sensing of a remote sensing image based on
  the priors of the reference image, IEEE Geoscience and Remote Sensing Letters
  12~(4) (2014) 736--740.

\bibitem{cai2020data}
J.-F. Cai, J.~K. Choi, K.~Wei, Data driven tight frame for compressed sensing
  {MRI} reconstruction via off-the-grid regularization, SIAM Journal on Imaging
  Sciences 13~(3) (2020) 1272--1301.

\bibitem{Yang2021}
G.~Yang, J.~Lv, Y.~Chen, J.~Huang, J.~Zhu, Generative Adversarial Network
  Powered Fast Magnetic Resonance Imaging---Comparative Study and New
  Perspectives, Springer International Publishing, Cham, 2022, pp. 305--339.

\bibitem{Lv2021_1}
J.~Lv, J.~Zhu, G.~Yang, Which {GAN}? {A} comparative study of generative
  adversarial network-based fast {MRI} reconstruction, Philosophical
  Transactions of the Royal Society A: Mathematical, Physical and Engineering
  Sciences 379 (2021) 20200203.
\newblock \href {https://doi.org/10.1098/rsta.2020.0203}
  {\path{doi:10.1098/rsta.2020.0203}}.

\bibitem{Shaul2020}
R.~Shaul, I.~David, O.~Shitrit, T.~R. Raviv, Subsampled brain {MRI}
  reconstruction by generative adversarial neural networks, Medical Image
  Analysis 65 (2020) 101747.
\newblock \href {https://doi.org/10.1016/j.media.2020.101747}
  {\path{doi:10.1016/j.media.2020.101747}}.

\bibitem{Huang2021}
J.~Huang, W.~Ding, J.~Lv, J.~Yang, H.~Dong, J.~{Del Ser}, J.~Xia, T.~Ren,
  S.~Wong, G.~Yang, Edge-enhanced dual discriminator generative adversarial
  network for fast {MRI} with parallel imaging using multi-view information,
  Applied Intelligence (2021).
\newblock \href {https://doi.org/10.1007/s10489-021-03092-w}
  {\path{doi:10.1007/s10489-021-03092-w}}.

\bibitem{Quan2018}
T.~M. Quan, T.~Nguyen-Duc, W.-K. Jeong, Compressed sensing {MRI} reconstruction
  using a generative adversarial network with a cyclic loss, IEEE Transactions
  on Medical Imaging 37 (2018) 1488--1497.
\newblock \href {https://doi.org/10.1109/TMI.2018.2820120}
  {\path{doi:10.1109/TMI.2018.2820120}}.

\bibitem{Ma2021}
Y.~Ma, J.~Liu, Y.~Liu, H.~Fu, Y.~Hu, J.~Cheng, H.~Qi, Y.~Wu, J.~Zhang, Y.~Zhao,
  Structure and illumination constrained gan for medical image enhancement,
  IEEE Transactions on Medical Imaging (2021) 1--1\href
  {https://doi.org/10.1109/TMI.2021.3101937}
  {\path{doi:10.1109/TMI.2021.3101937}}.

\bibitem{Arjovsky2017}
M.~Arjovsky, S.~Chintala, L.~Bottou, {W}asserstein generative adversarial
  networks, in: Proceedings of the 34th International Conference on Machine
  Learning, Vol.~70 of Proceedings of Machine Learning Research, PMLR, 2017,
  pp. 214--223.

\bibitem{Guo2020}
Y.~Guo, C.~Wang, H.~Zhang, G.~Yang, Deep attentive wasserstein generative
  adversarial networks for {MRI} reconstruction with recurrent
  context-awareness, in: Medical Image Computing and Computer Assisted
  Intervention -- MICCAI 2020, Springer International Publishing, Cham, 2020,
  pp. 167--177.

\bibitem{Jiang2021}
M.~Jiang, M.~Zhi, L.~Wei, X.~Yang, J.~Zhang, Y.~Li, P.~Wang, J.~Huang, G.~Yang,
  {FA-GAN}: Fused attentive generative adversarial networks for {MRI} image
  super-resolution, Computerized Medical Imaging and Graphics 92 (2021) 101969.
\newblock \href {https://doi.org/10.1016/j.compmedimag.2021.101969}
  {\path{doi:10.1016/j.compmedimag.2021.101969}}.

\bibitem{Ronneberger2015}
O.~Ronneberger, P.~Fischer, T.~Brox, {U-Net}: Convolutional networks for
  biomedical image segmentation, in: Medical Image Computing and
  Computer-Assisted Intervention -- MICCAI 2015, Springer International
  Publishing, Cham, 2015, pp. 234--241.

\bibitem{Lv2021_3}
J.~Lv, G.~Li, X.~Tong, W.~Chen, J.~Huang, C.~Wang, G.~Yang, Transfer learning
  enhanced generative adversarial networks for multi-channel {MRI}
  reconstruction, Computers in Biology and Medicine 134 (2021) 104504.
\newblock \href {https://doi.org/10.1016/j.compbiomed.2021.104504}
  {\path{doi:10.1016/j.compbiomed.2021.104504}}.

\bibitem{Uecker2014}
M.~Uecker, P.~Lai, M.~J. Murphy, P.~Virtue, M.~Elad, J.~M. Pauly, S.~S.
  Vasanawala, M.~Lustig, {ESPIRiT}-an eigenvalue approach to autocalibrating
  parallel {MRI}: Where {SENSE} meets {GRAPPA}, Magnetic Resonance in Medicine
  71~(3) (2014) 990--1001.

\bibitem{Lai2019}
W.-S. Lai, J.-B. Huang, N.~Ahuja, M.-H. Yang, Fast and accurate image
  {Super-Resolution} with deep laplacian pyramid networks, IEEE Transactions on
  Pattern Analysis and Machine Intelligence 41~(11) (2019) 2599--2613.
\newblock \href {https://doi.org/10.1109/TPAMI.2018.2865304}
  {\path{doi:10.1109/TPAMI.2018.2865304}}.

\bibitem{Souza2018}
R.~Souza, O.~Lucena, J.~Garrafa, D.~Gobbi, M.~Saluzzi, S.~Appenzeller,
  L.~Rittner, R.~Frayne, R.~Lotufo, An open, multi-vendor, multi-field-strength
  brain {MR} dataset and analysis of publicly available skull stripping methods
  agreement, NeuroImage 170 (2018) 482--494, segmenting the Brain.
\newblock \href
  {https://doi.org/https://doi.org/10.1016/j.neuroimage.2017.08.021}
  {\path{doi:https://doi.org/10.1016/j.neuroimage.2017.08.021}}.

\bibitem{BraTS17_1}
B.~H. Menze, A.~Jakab, S.~Bauer, J.~Kalpathy-Cramer, K.~Farahani, J.~Kirby,
  Y.~Burren, N.~Porz, J.~Slotboom, R.~Wiest, et~al., The multimodal brain tumor
  image segmentation benchmark ({BRATS}), IEEE transactions on medical imaging
  34~(10) (2014) 1993--2024.

\bibitem{BraTS17_2}
S.~Bakas, H.~Akbari, A.~Sotiras, M.~Bilello, M.~Rozycki, J.~S. Kirby, J.~B.
  Freymann, K.~Farahani, C.~Davatzikos, Advancing the cancer genome atlas
  glioma {MRI} collections with expert segmentation labels and radiomic
  features, Scientific data 4~(1) (2017) 1--13.

\bibitem{BraTS17_3}
S.~{Bakas}, M.~{Reyes}, A.~{Jakab}, S.~{Bauer}, M.~{Rempfler}, A.~{Crimi},
  R.~{Takeshi Shinohara}, C.~{Berger}, S.~M. {Ha}, M.~{Rozycki}, M.~{Prastawa},
  E.~{Alberts}, J.~{Lipkova}, J.~{Freymann}, J.~{Kirby}, M.~{Bilello},
  H.~{Fathallah-Shaykh}, R.~{Wiest}, J.~{Kirschke}, B.~{Wiestler}, R.~{Colen},
  A.~{Kotrotsou}, P.~{Lamontagne}, D.~{Marcus}, M.~{Milchenko}, A.~{Nazeri},
  M.-A. {Weber}, A.~{Mahajan}, U.~{Baid}, E.~{Gerstner}, D.~{Kwon},
  G.~{Acharya}, M.~{Agarwal}, M.~{Alam}, A.~{Albiol}, A.~{Albiol}, F.~J.
  {Albiol}, V.~{Alex}, N.~{Allinson}, P.~H.~A. {Amorim}, A.~{Amrutkar},
  G.~{Anand}, S.~{Andermatt}, T.~{Arbel}, P.~{Arbelaez}, A.~{Avery},
  M.~{Azmat}, B.~{Pranjal}, W.~{Bai}, S.~{Banerjee}, B.~{Barth},
  T.~{Batchelder}, K.~{Batmanghelich}, E.~{Battistella}, A.~{Beers},
  M.~{Belyaev}, M.~{Bendszus}, E.~{Benson}, J.~{Bernal}, H.~{Nagaraja Bharath},
  G.~{Biros}, S.~{Bisdas}, J.~{Brown}, M.~{Cabezas}, S.~{Cao}, J.~M. {Cardoso},
  E.~N. {Carver}, A.~{Casamitjana}, L.~{Silvana Castillo}, M.~{Cat{\`a}},
  P.~{Cattin}, A.~{Cerigues}, V.~S. {Chagas}, S.~{Chandra}, Y.-J. {Chang},
  S.~{Chang}, K.~{Chang}, J.~{Chazalon}, S.~{Chen}, W.~{Chen}, J.~W. {Chen},
  Z.~{Chen}, K.~{Cheng}, A.~R. {Choudhury}, R.~{Chylla}, A.~{Cl{\'e}rigues},
  S.~{Colleman}, R.~{German Rodriguez Colmeiro}, M.~{Combalia}, A.~{Costa},
  X.~{Cui}, Z.~{Dai}, L.~{Dai}, L.~A. {Daza}, E.~{Deutsch}, C.~{Ding},
  C.~{Dong}, S.~{Dong}, W.~{Dudzik}, Z.~{Eaton-Rosen}, G.~{Egan},
  G.~{Escudero}, T.~{Estienne}, R.~{Everson}, J.~{Fabrizio}, Y.~{Fan},
  L.~{Fang}, X.~{Feng}, E.~{Ferrante}, L.~{Fidon}, M.~{Fischer}, A.~P.
  {French}, N.~{Fridman}, H.~{Fu}, D.~{Fuentes}, Y.~{Gao}, E.~{Gates},
  D.~{Gering}, A.~{Gholami}, W.~{Gierke}, B.~{Glocker}, M.~{Gong},
  S.~{Gonz{\'a}lez-Vill{\'a}}, T.~{Grosges}, Y.~{Guan}, S.~{Guo}, S.~{Gupta},
  W.-S. {Han}, I.~S. {Han}, K.~{Harmuth}, H.~{He},
  A.~{Hern{\'a}ndez-Sabat{\'e}}, E.~{Herrmann}, N.~{Himthani}, W.~{Hsu},
  C.~{Hsu}, X.~{Hu}, X.~{Hu}, Y.~{Hu}, Y.~{Hu}, R.~{Hua}, T.-Y. {Huang},
  W.~{Huang}, S.~{Van Huffel}, Q.~{Huo}, V.~{HV}, K.~M. {Iftekharuddin},
  F.~{Isensee}, M.~{Islam}, A.~S. {Jackson}, S.~R. {Jambawalikar}, A.~{Jesson},
  W.~{Jian}, P.~{Jin}, V.~J.~M. {Jose}, A.~{Jungo}, B.~{Kainz}, K.~{Kamnitsas},
  P.-Y. {Kao}, A.~{Karnawat}, T.~{Kellermeier}, A.~{Kermi}, K.~{Keutzer},
  M.~{Tarek Khadir}, M.~{Khened}, P.~{Kickingereder}, G.~{Kim}, N.~{King},
  H.~{Knapp}, U.~{Knecht}, L.~{Kohli}, D.~{Kong}, X.~{Kong}, S.~{Koppers},
  A.~{Kori}, G.~{Krishnamurthi}, E.~{Krivov}, P.~{Kumar}, K.~{Kushibar},
  D.~{Lachinov}, T.~{Lambrou}, J.~{Lee}, C.~{Lee}, Y.~{Lee}, M.~{Lee},
  S.~{Lefkovits}, L.~{Lefkovits}, J.~{Levitt}, T.~{Li}, H.~{Li}, W.~{Li},
  H.~{Li}, X.~{Li}, Y.~{Li}, H.~{Li}, Z.~{Li}, X.~{Li}, Z.~{Li}, X.~{Li},
  W.~{Li}, Z.-S. {Lin}, F.~{Lin}, P.~{Lio}, C.~{Liu}, B.~{Liu}, X.~{Liu},
  M.~{Liu}, J.~{Liu}, L.~{Liu}, X.~{Llado}, M.~{Moreno Lopez}, P.~{Ribalta
  Lorenzo}, Z.~{Lu}, L.~{Luo}, Z.~{Luo}, J.~{Ma}, K.~{Ma}, T.~{Mackie},
  A.~{Madabushi}, I.~{Mahmoudi}, K.~H. {Maier-Hein}, P.~{Maji}, C.~{Mammen},
  A.~{Mang}, B.~S. {Manjunath}, M.~{Marcinkiewicz}, S.~{McDonagh},
  S.~{McKenna}, R.~{McKinley}, M.~{Mehl}, S.~{Mehta}, R.~{Mehta}, R.~{Meier},
  C.~{Meinel}, D.~{Merhof}, C.~{Meyer}, R.~{Miller}, S.~{Mitra}, A.~{Moiyadi},
  D.~{Molina-Garcia}, M.~A.~B. {Monteiro}, G.~{Mrukwa}, A.~{Myronenko},
  J.~{Nalepa}, T.~{Ngo}, D.~{Nie}, H.~{Ning}, C.~{Niu}, N.~K. {Nuechterlein},
  E.~{Oermann}, A.~{Oliveira}, D.~D.~C. {Oliveira}, A.~{Oliver}, A.~F.~I.
  {Osman}, Y.-N. {Ou}, S.~{Ourselin}, N.~{Paragios}, M.~S. {Park},
  B.~{Paschke}, J.~G. {Pauloski}, K.~{Pawar}, N.~{Pawlowski}, L.~{Pei},
  S.~{Peng}, S.~M. {Pereira}, J.~{Perez-Beteta}, V.~M. {Perez-Garcia},
  S.~{Pezold}, B.~{Pham}, A.~{Phophalia}, G.~{Piella}, G.~N. {Pillai},
  M.~{Piraud}, M.~{Pisov}, A.~{Popli}, M.~P. {Pound}, R.~{Pourreza},
  P.~{Prasanna}, V.~{Prkovska}, T.~P. {Pridmore}, S.~{Puch},
  {\'E}.~{Puybareau}, B.~{Qian}, X.~{Qiao}, M.~{Rajchl}, S.~{Rane},
  M.~{Rebsamen}, H.~{Ren}, X.~{Ren}, K.~{Revanuru}, M.~{Rezaei}, O.~{Rippel},
  L.~C. {Rivera}, C.~{Robert}, B.~{Rosen}, D.~{Rueckert}, M.~{Safwan},
  M.~{Salem}, J.~{Salvi}, I.~{Sanchez}, I.~{S{\'a}nchez}, H.~M. {Santos},
  E.~{Sartor}, D.~{Schellingerhout}, K.~{Scheufele}, M.~R. {Scott}, A.~A.
  {Scussel}, S.~{Sedlar}, J.~P. {Serrano-Rubio}, N.~J. {Shah}, N.~{Shah},
  M.~{Shaikh}, B.~U. {Shankar}, Z.~{Shboul}, H.~{Shen}, D.~{Shen}, L.~{Shen},
  H.~{Shen}, V.~{Shenoy}, F.~{Shi}, H.~E. {Shin}, H.~{Shu}, D.~{Sima},
  M.~{Sinclair}, O.~{Smedby}, J.~M. {Snyder}, M.~{Soltaninejad}, G.~{Song},
  M.~{Soni}, J.~{Stawiaski}, S.~{Subramanian}, L.~{Sun}, R.~{Sun}, J.~{Sun},
  K.~{Sun}, Y.~{Sun}, G.~{Sun}, S.~{Sun}, Y.~R. {Suter}, L.~{Szilagyi},
  S.~{Talbar}, D.~{Tao}, D.~{Tao}, Z.~{Teng}, S.~{Thakur}, M.~H. {Thakur},
  S.~{Tharakan}, P.~{Tiwari}, G.~{Tochon}, T.~{Tran}, Y.~M. {Tsai}, K.-L.
  {Tseng}, T.~A. {Tuan}, V.~{Turlapov}, N.~{Tustison}, M.~{Vakalopoulou},
  S.~{Valverde}, R.~{Vanguri}, E.~{Vasiliev}, J.~{Ventura}, L.~{Vera},
  T.~{Vercauteren}, C.~A. {Verrastro}, L.~{Vidyaratne}, V.~{Vilaplana},
  A.~{Vivekanandan}, G.~{Wang}, Q.~{Wang}, C.~J. {Wang}, W.~{Wang}, D.~{Wang},
  R.~{Wang}, Y.~{Wang}, C.~{Wang}, G.~{Wang}, N.~{Wen}, X.~{Wen},
  L.~{Weninger}, W.~{Wick}, S.~{Wu}, Q.~{Wu}, Y.~{Wu}, Y.~{Xia}, Y.~{Xu},
  X.~{Xu}, P.~{Xu}, T.-L. {Yang}, X.~{Yang}, H.-Y. {Yang}, J.~{Yang},
  H.~{Yang}, G.~{Yang}, H.~{Yao}, X.~{Ye}, C.~{Yin}, B.~{Young-Moxon}, J.~{Yu},
  X.~{Yue}, S.~{Zhang}, A.~{Zhang}, K.~{Zhang}, X.~{Zhang}, L.~{Zhang},
  X.~{Zhang}, Y.~{Zhang}, L.~{Zhang}, J.~{Zhang}, X.~{Zhang}, T.~{Zhang},
  S.~{Zhao}, Y.~{Zhao}, X.~{Zhao}, L.~{Zhao}, Y.~{Zheng}, L.~{Zhong},
  C.~{Zhou}, X.~{Zhou}, F.~{Zhou}, H.~{Zhu}, J.~{Zhu}, Y.~{Zhuge}, W.~{Zong},
  J.~{Kalpathy-Cramer}, K.~{Farahani}, C.~{Davatzikos}, K.~{van Leemput},
  B.~{Menze}, Identifying the best machine learning algorithms for brain tumor
  segmentation, progression assessment, and overall survival prediction in the
  {BRATS} challenge, arXiv e-prints (2018) arXiv:1811.02629.

\bibitem{Heusel2017}
M.~Heusel, H.~Ramsauer, T.~Unterthiner, B.~Nessler, S.~Hochreiter, Gans trained
  by a two time-scale update rule converge to a local nash equilibrium,
  Advances in neural information processing systems 30 (2017).

\bibitem{Zhang2018}
R.~Zhang, P.~Isola, A.~A. Efros, E.~Shechtman, O.~Wang, The unreasonable
  effectiveness of deep features as a perceptual metric, in: Proceedings of the
  IEEE Conference on Computer Vision and Pattern Recognition (CVPR), 2018.

\bibitem{Hansen2015}
M.~S. Hansen, P.~Kellman, Image reconstruction: An overview for clinicians,
  Journal of Magnetic Resonance Imaging 41~(3) (2015) 573--585.
\newblock \href {https://doi.org/https://doi.org/10.1002/jmri.24687}
  {\path{doi:https://doi.org/10.1002/jmri.24687}}.

\bibitem{Noori2019}
M.~Noori, A.~Bahri, K.~Mohammadi, Attention-guided version of {2D} {UNet} for
  automatic brain tumor segmentation, in: 2019 9th International Conference on
  Computer and Knowledge Engineering (ICCKE), 2019, pp. 269--275.
\newblock \href {https://doi.org/10.1109/ICCKE48569.2019.8964956}
  {\path{doi:10.1109/ICCKE48569.2019.8964956}}.

\bibitem{Hu2017}
J.~Hu, L.~Shen, G.~Sun, Squeeze-and-excitation networks, in: Proceedings of the
  IEEE Conference on Computer Vision and Pattern Recognition (CVPR), 2018.

\end{thebibliography}


%% \begin{thebibliography}{00}


%% \bibitem{}

%% \end{thebibliography}
\end{document}